\PassOptionsToPackage{dvipsnames,table}{xcolor}
\documentclass[acmsmall,screen]{acmart}
\usepackage{multirow}
\usepackage{booktabs}
\usepackage{graphicx}
\usepackage{array}
\usepackage[most]{tcolorbox}
\usepackage{colortbl}
\usepackage{amsmath}
\usepackage{xspace}
\usepackage{enumitem}
\usepackage{hyperref}
\usepackage{listings}
\usepackage{pifont}
\usepackage{soul}
\usepackage[dvipsnames]{xcolor}
\usepackage{url}
\usepackage[ruled,vlined]{algorithm2e}
\usepackage{tabularx}
\usepackage{caption}
\usepackage{fontawesome5}
\usepackage{balance}
\usepackage{seqsplit} 
\usepackage{graphicx}
\usepackage{subcaption}
\usepackage{wrapfig}
\usepackage{xcolor}

\newtcolorbox{finding}[1]{
  colframe=green!40!black, 
  colback=green!5!white,
  coltitle=white,
  fonttitle=\bfseries,
  title={\faLightbulb\hspace{0.5em}\textbf{#1}},
  sharp corners,
  boxrule=1pt,
  boxsep=1mm, 
  left=2mm,   
  right=2mm,  
  top=1mm,    
  bottom=1mm  
}

\tcbuselibrary{skins}
\newtcolorbox{experience}{
    colback=white,
    colframe=white,
    boxrule=0pt,
    leftrule=2pt,
    colframe=gray!50,
    left=8pt,
    top=1pt,
    bottom=1pt,
    enhanced jigsaw,
    sharp corners,
    boxsep=0pt,
    arc=0pt,
    outer arc=0pt,
    fontupper=\small
}

\tcbset{
  promptbox/.style={
    enhanced,
    colback=gray!5,
    colframe=gray!50,
    boxrule=0.5pt,
    arc=2mm,
    outer arc=2mm,
    fonttitle=\bfseries,
    coltitle=black,
    title={#1},
    sharp corners=south,
    breakable,
    listing only,
    listing options={basicstyle=\ttfamily\footnotesize, breaklines=true}
  }
}

\usepackage[utf8]{inputenc}   
\usepackage[T1]{fontenc}
\usepackage{upquote}
\usepackage{listings}

\newcommand{\HLred}{\color{red}\bfseries}
\newcommand{\HLgreen}{\color{ForestGreen}\bfseries}

\definecolor{keycolor}{RGB}{176,114,25}
\definecolor{valuecolor}{RGB}{0,128,0}

\author{Silin Chen}
\affiliation{%
  \institution{Shanghai Jiao Tong University}
  \country{China}}
\email{cslsolow@gmail.com}

\author{Shaoxin Lin}
\affiliation{%
  \institution{Huawei}
  \country{China}}
\email{2120200411@mail.nankai.edu.cn}

\author{Yuling Shi}
\affiliation{%
  \institution{Shanghai Jiao Tong University}
  \country{China}}
\email{yuling.shi@sjtu.edu.cn}

\author{Heng Lian}
\affiliation{%
  \institution{Xidian University}
  \country{China}}
\email{lianheng23@163.com}

\author{Xiaodong Gu\textsuperscript{\textdagger}}
\affiliation{%
  \institution{Shanghai Jiao Tong University}
  \country{China}}
\email{xiaodong.gu@sjtu.edu.cn}

\author{Longfei Yun}
\affiliation{%
  \institution{UC San Diego}
  \country{United States}}
\email{loyun@ucsd.edu}

\author{Dong Chen}
\affiliation{%
  \institution{Huawei}
  \country{China}}
\email{jameschennerd@gmail.com}

\author{Lin Cao}
\affiliation{%
  \institution{Huawei}
  \country{China}}
\email{robin_a@126.com}

\author{Jiyang Liu}
\affiliation{%
  \institution{Huawei}
  \country{China}}
\email{jiyangliu@berkeley.edu}

\author{Nu Xia}
\affiliation{%
  \institution{Huawei}
  \country{China}}
\email{xianu@huawei.com}

\author{Qianxiang Wang}
\affiliation{%
  \institution{Huawei}
  \country{China}}
\email{wangqianxiang@huawei.com}

\newcommand{\approach}{{SWE-Exp}\xspace}

\newcommand{\lparagraph}[1]{\textbf{#1}~}

\title{\approach: Experience-Driven Software Issue Resolution}

\begin{document}

\begin{abstract}
    Recent advances in large language model (LLM) agents have shown remarkable progress in software issue resolution, leveraging advanced techniques such as multi-agent collaboration and Monte Carlo Tree Search (MCTS). 
    However, current agents act as memoryless explorers—treating each problem separately without retaining or reusing knowledge from previous repair experiences. 
    This leads to redundant exploration of failed trajectories and missed chances to adapt successful issue resolution methods to similar problems.
    To address this problem, we introduce \approach, an experience-enhanced approach that distills concise and actionable experience from prior agent trajectories, enabling continuous learning across issues. 
    Our method introduces a multi-faceted experience bank that captures both successful and failed repair attempts. Specifically, it extracts reusable issue resolution knowledge at different levels—from high-level problem comprehension to specific code changes. 
    Experiments show that \approach achieves a Pass@1 resolution rate of 73.0\% on SWE-Bench Verified using the state-of-the-art LLM Claude 4 Sonnet, significantly outperforming prior results under other agent frameworks.
    Our approach establishes a new paradigm in which automated software engineering agents systematically accumulate and leverage repair expertise, fundamentally shifting from trial-and-error exploration to strategic, experience-driven issue resolution.
\end{abstract}

\maketitle

\section{Introduction}
 
Software issue resolution, which aims to automatically localize and fix faults across interdependent source files, represents one of the most challenging tasks in automated software engineering~\cite{jimenez2024swebench,chen2025unveilinga,shao2025llms,shi2024between, shi2025longcodezip, peng2025swe}.
With the introduction of SWE-bench~\cite{jimenez2024swebench}—the standard benchmark for evaluating automated program repair (APR) on real-world GitHub issues—researchers have developed a diverse range of techniques to tackle this challenge~\cite{wang2026swe,ma2025improving,chen2025locagent,chen2024coder,antoniades2024swesearch}. SWE-bench provides a comprehensive evaluation framework by pairing real-world GitHub issues with their full repository-level contexts, enabling rigorous assessment of repair methods in realistic, complex software environments.


Recently, the emergence of LLMs and multi-agent techniques has significantly advanced the task of automated issue resolution. Agent-based approaches equip LLMs with external tools for code navigation, editing, and testing, enabling them to iteratively explore and refine potential solutions~\cite{yang2024sweagenta,chen2024coder,antoniades2024swesearch}. Building on this foundation, MCTS-based systems further enhance agent capabilities by guiding exploration in a more systematic and goal-directed manner~\cite{antoniades2024swesearch}, improving the efficiency and completeness of the search process. 

Despite remarkable progress in the issue-solving rate, current approaches suffer from a fundamental limitation: agents operate as memoryless explorers, treating each issue in isolation and failing to leverage insights from previous repair attempts~\cite{cuadron2025danger}. This limitation creates three critical inefficiencies: 1) Redundant exploration: agents often retry ineffective trajectories across similar issues, expending computational efforts on issue resolution strategies that have proven unsuccessful in similar contexts~\cite{antoniades2024swesearch, lin2024llms}; 2) Inability of knowledge transfer: agents often discard valuable insights from successful resolution trajectories, including effective issue resolution workflows, code patterns, and contextual factors influencing patch quality after each session~\cite{zhao2024expelc,xia2024agentless, yang2024sweagenta}; and 3) Lack of strategic evolution: without systematic experience accumulation, agents are unable to develop increasingly refined issue resolution strategies or compound expertise over time. As a result, they struggle to adapt to novel or evolving issues, particularly those that are specific to individual repositories~\cite{robeyns2025selfimproving, lin2024llms}. 


\begin{figure*}[t]
    \centering
    \includegraphics[width=\textwidth]{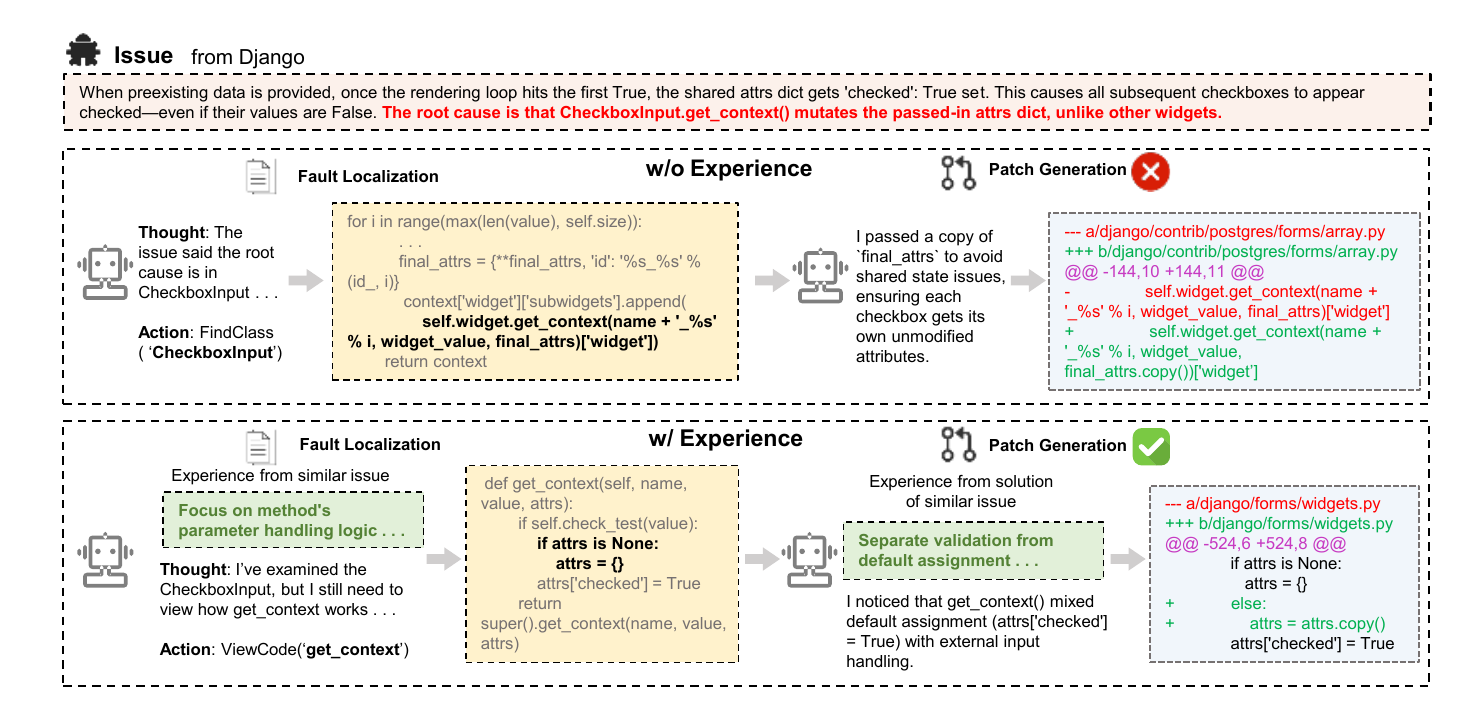}
    \caption{A motivating example of experience-guided approach on instance \textit{django-11964}.}
    \label{fig:motivation}
\end{figure*}

To address these challenges, we propose \approach, a novel experience-enhanced approach that transforms issue resolution from isolated, stateless problem-solving into a continuous learning process. Unlike previous approaches that only utilize internal knowledge~\cite{wei2025swerl,yang2024sweagenta}, \approach distills structured experiences from prior resolution attempts, and leverages such accumulated knowledge to guide future repair attempts. 
\approach maintains an evolving experience bank that encodes knowledge across three facets: trajectory-guided problem understanding, fault localization patterns, and modification strategies. When encountering a new issue, the system retrieves relevant experiences and distills them into actionable guidance. To operationalize this knowledge more efficiently, we employ a dual-agent architecture, where an \textit{Instructor} agent formulates high-level strategies and an \textit{Assistant} agent executes low-level operations.
By maintaining an evolving multi-facet experience bank, our approach avoids redundant exploration and leverages past knowledge more effectively.

We empirically validate \approach on SWE-Bench Verified~\cite{jimenez2024swebench}, a program repair benchmark of 500 human-verified GitHub issues. Our experiments demonstrate that \approach achieves a Pass@1 resolution rate of 73\% using Claude 4 Sonnet, significantly outperforming prior results under other agent frameworks. Furthermore, the comprehension and modification capabilities distilled by \approach independently contribute to performance improvements, with their combination yielding the most significant gains. These findings highlight the broader potential of structured reasoning and iterative refinement within agent frameworks, enabling more effective problem-solving in software engineering tasks.

Our main contributions can be summarized as follows:

\begin{itemize}
    \item We present a novel framework that systematically captures and manages experiences from agent trajectories at multiple facets, enabling systematic collection of repair knowledge across different issue contexts.
    \item We propose an experience-driven system that uses historical knowledge through dynamic retrieval to improve fault-localization accuracy and patch quality, transforming repository-level issue resolution from isolated problem-solving into continuous learning.
    \item Experimental validation showing that \approach achieves state-of-the-art resolution rate on open-source agent frameworks.
\end{itemize}

\section{Motivation}
Recent advances in MCTS approaches for repository-level issue resolution have shown significant improvements in exploring solution spaces systematically~\cite{antoniades2024swesearch}. These methods enable agents to backtrack and explore alternative solutions through strategic search tree expansion, addressing limitations of linear sequential processes. However, even with advanced search strategies, current agents remain fundamentally limited by their inability to learn from accumulated issue resolution experiences across different issues. 

To illustrate this limitation, let us consider a concrete issue resolution scenario from the Django codebase involving a composite widget composed of multiple checkboxes. The issue manifests as all checkbox widgets appearing checked regardless of their actual values. This is caused by the \texttt{Checkbox\allowbreak Input} widget modifying a shared \texttt{attrs} dictionary in place. Figure~\ref{fig:motivation} demonstrates how agents approach this problem with and without historical experience.

When operating without experience, the agent focuses on the surface-level symptoms mentioned in the issue description. This leads to a patch that creates a copy of \texttt{final\_attrs} within the context rendering method of the composite widget. While this appears to address the immediate problem, it represents a narrow, symptom-focused solution that fails to address the root cause: the \texttt{Checkbox\allowbreak Input} widget's fundamental design flaw of modifying its input parameters. This approach has two critical limitations. First, it only addresses the specific context for the particular class, leaving the underlying issue unresolved for other potential widget combinations. Second, it treats the symptom rather than the disease, creating a fragile fix that may not prevent similar issues in future scenarios.

In contrast, an agent equipped with relevant experience demonstrates more strategic and insightful reasoning. Prior experience directs the agent to examine not just the surface behavior but the deeper mechanism—specifically, how method parameters are handled and potentially mutated during execution. This perspective leads the agent to scrutinize the implementation of the \texttt{Checkbox\allowbreak Input.\allowbreak get\_context()} method, where the root cause resides. Drawing on its accumulated repair knowledge, the agent applies a principled fix: modifying the method to create a defensive copy of the \texttt{attrs} dictionary before performing any changes. This strategy reflects a key insight from past experience—that default assignments need to be verified and handled separately to avoid unintended side effects. Compared to symptom-level patches, this solution is significantly more robust, as it eliminates the underlying design flaw and ensures correctness across diverse usage contexts. It exemplifies how experiential knowledge enables agents to fix problems at their source, rather than applying narrow, reactive workarounds.

\begin{figure*}[t]
  \centering
  \includegraphics[width=\textwidth]{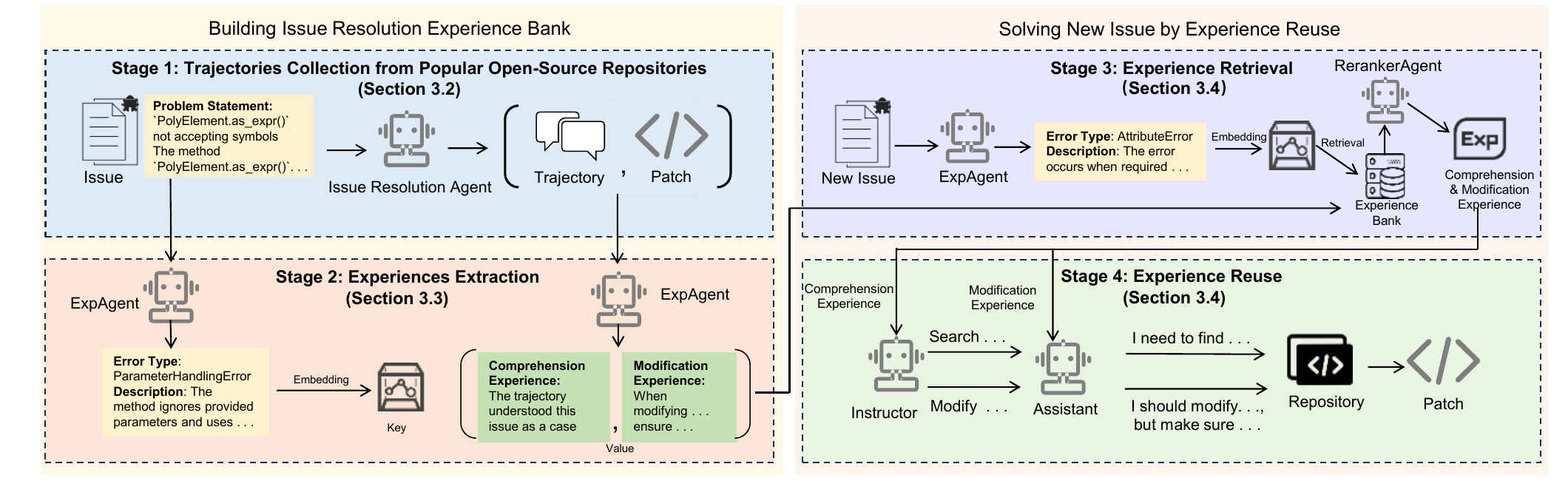}
  \caption{The framework of \approach.}
  \label{fig:framework}
\end{figure*}

This example reveals a critical insight: the difference between surface-level symptom fixing and deep root-cause resolution. Without systematic experience accumulation, agents repeatedly engage in reactive issue resolution that addresses immediate symptoms without understanding underlying patterns. Experience-guided agents, however, develop pattern recognition capabilities that enable them to identify and resolve fundamental issues. This leads to more robust and generalizable solutions. This motivates our design of \approach, which transforms repository-level issue resolution from isolated symptom-focused issue resolution into a systematic, knowledge-driven process that accumulates and uses repair expertise across similar contexts.

\section{Methodology}
\label{sec:approach}

\subsection{Conceptual Framework}
\label{sec:conceptual}

To formalize our approach, consider a standard agent that resolves an issue $p$ by generating a reasoning and action trajectory $\tau = (s_0, a_0, s_1, a_1, \ldots)$, where $s_t$ denotes the repository state at time step $t$ and $a_t$ is the corresponding action taken. Existing methods typically treat each trajectory in the agent’s history $\mathcal{H} = \{\tau_1, \tau_2, \ldots\}$ as independent and do not exploit prior experiences during new problem solving. In contrast, our approach introduces an experience bank 
$\mathcal{B}_{\text{exp}}$ which systematically accumulates distilled knowledge from the agent’s past resolution trajectories. When faced with a new issue instance $p'$, the agent queries the memory to retrieve relevant experiences $E' = {Retrieve}(p', \mathcal{B}_{\text{exp}})$. The subsequent search for a new trajectory $\tau'$ is then conditioned on the retrieved experience set $E'$. This transforms the agent’s behavior from blind exploration to experience-guided reasoning, allowing it to leverage prior insights and quickly converge toward final solutions.


The framework follows a systematic four-stage pipeline, as illustrated in Figures~\ref{fig:framework}: First, the system collects repair trajectories from both successful and failed issue resolution attempts across diverse repository contexts (Section~\ref{sec:trajectories}). Second, an offline experience extraction process transforms these raw trajectories into structured, multi-faceted knowledge at different abstraction levels (Section~\ref{sec:extraction}). Finally, when facing new issues, the system retrieves and adapts relevant historical experiences to provide targeted guidance, and experience-informed agents execute the issue resolution process through strategic planning and tactical implementation (Section~\ref{sec:guidance}).

\subsection{Trajectories Collection}
\label{sec:trajectories}

Our approach begins with the systematic collection of repair trajectories as the source for distilling experience. 

\lparagraph{Structured Trajectory Representation.} Each agent repair attempt, successful or not, is represented as a sequence of tuples $\langle(d_t, a_t, s_{t+1}, f_t)\rangle_{t=0..N}$ where $d_t$ represents high-level directives, $a_t$ denotes specific actions taken, $s_{t+1}$ captures the resulting repository state, and $f_t$ denotes environment feedback. This structural representation ensures that both successful issue resolution workflows and failure patterns are preserved for later analysis.


\lparagraph{Success Or Failure Annotation.} Each trajectory is annotated with success (i.e., producing correct patches validated against ground truth) or failure. For failure trajectories, additional metadata is appended to capture the failure cause—whether due to incorrect localization, flawed modification strategy, or insufficient problem comprehension. Such a categorization enables targeted experience extraction for both positive and negative patterns.

\subsection{Experiences Extraction}
\label{sec:extraction}

Raw trajectories, while comprehensive, are often too lengthy, noisy, and problem-specific to be directly reused. Our next stage aims to transform them into structured, reusable knowledge. An \textit{Experiencer} agent is instructed to extract experiences from both success and failure trajectories. 

\subsubsection{Experience Representation.} 
We define \textit{experience} as the generalized and transferable thinking patterns extracted from an agent's past issue-solving process. It consists of two key components:
 \begin{itemize}[leftmargin=0.5cm]
     \item \textbf{Perspective}: the agent's abstract understanding of the problem.
     \item \textbf{Modification}: the generalized strategy used to address the issue.
 \end{itemize}
Formally, each experience is represented as a dictionary, where the key denotes the perspective and the value represents the corresponding modifications. For example, 
\begin{lstlisting}
  (*@\textcolor{keycolor}{"perspective"}@*): (*@\textcolor{valuecolor}{"The trajectory understood this issue as a deprecation of legacy behavior that was no longer necessary due to improvements in the system's handling of structured data. The perspective focused on transitioning users smoothly from an old implementation to a more direct approach."}@*),
  (*@\textcolor{keycolor}{"modification"}@*): [ 
    (*@\textcolor{valuecolor}{"When deprecating functionality, it's important to first add a warning before removing the feature, giving users time to adapt their code."}@*),
    (*@\textcolor{valuecolor}{"Removing automatic type conversions can simplify code and make behavior more predictable, but requires careful consideration of backward compatibility."}@*)
  ]
\end{lstlisting}

\subsubsection{Offline Embedding and Storage}

To facilitate efficient retrieval during issue resolution, all extracted experiences are embedded and stored in a vector database, which we refer to as the \textit{Experience Bank}. During the offline embedding process, each experience is indexed by two metadata attributes: 
\begin{itemize}[leftmargin=0.5cm]
    \item \textbf{Issue Type}: A generalized, descriptive label inferred by the agent based on the issue, such as the \texttt{AttributeError} and the \texttt{VariableReferenceError}.
    \item \textbf{Description}: A generalized explanation generated by the agent, describing the typical conditions and scenarios in which this type of error arises.
\end{itemize}
For instance, 
\begin{lstlisting}
      (*@\textcolor{keycolor}{"Issue"}@*): (*@\textcolor{valuecolor}{sphinx-doc\_\_sphinx-8638}@*),
      (*@\textcolor{keycolor}{"issue\_type"}@*): (*@\textcolor{valuecolor}{"VariableReferenceError"}@*),
      (*@\textcolor{keycolor}{"description"}@*): (*@\textcolor{valuecolor}{"Occurs when instance variables are incorrectly linked to other variables of the same name in the project, leading to unintended documentation references."}@*)
\end{lstlisting}

These attributes are encoded into dense vectors using a pre-trained embedding model. The resulting embeddings are stored in the Experience Bank, enabling semantic similarity search and retrieval during online resolution tasks.

\subsubsection{Multi-facet Categorization}
To support efficient and context-aware reuse, extracted experiences are organized into two categories, each reflecting a distinct facet of abstraction:

\lparagraph{Comprehension Experiences} These experiences capture how past issues were interpreted and reasoned about at a conceptual level. They encode general reasoning patterns for issue understanding, such as identifying key symptoms, forming diagnostic hypotheses, and leveraging contextual or structural cues to guide early-stage exploration. For instance, 
\begin{experience}
"The issue was fundamentally about how Sphinx handles variable linking in documentation, specifically the automatic linking of similarly named variables across different contexts (instance vs global). The golden patch reveals that the solution was to modify the role assignment for variable documentation fields rather than changing the fuzzy matching logic."
\end{experience}
\begin{experience}
"The core misunderstanding was focusing on the cross-referencing behavior (find\_obj method) rather than examining how variable documentation fields are processed and assigned roles in the Python domain."
\end{experience}
Comprehensive experiences inform how agents interpret and navigate unfamiliar issues, helping the agent prioritize relevant information and narrow the search space effectively.

\lparagraph{Modification Experiences} These experiences encode generalized strategies for code modification based on prior patches. They include insights into how responsibility was assigned to specific code regions, which behavioral contracts were violated, and how safety, scope, and potential side effects were assessed and managed.
For instance, 
\begin{experience}
    "When modifying a method that accepts optional parameters, ensure the logic properly handles both the presence and absence of these parameters without accidentally overriding valid inputs.",
\end{experience}
\begin{experience}
    "For methods that validate input parameters, structure the validation logic to clearly separate the validation step from the default value assignment to prevent unintended behavior."
\end{experience}
Modification experiences guide not only the structure of the fix but also the underlying reasoning and design choices that informed the patch.

The multi-faceted \textit{Experience Bank} serves as an external knowledge base that supports decision-making across the agent’s debugging pipeline. During issue resolution, relevant experiences are retrieved and used to guide both high-level diagnostic reasoning and low-level code editing. This enables agents to shift from trial-and-error exploration to strategic, experience-driven behavior.

\subsection{Experiences Reuse}
\label{sec:guidance}

Equipped with the experience bank, the agent is now ready to execute the core issue resolution task. This process unfolds across a standard three-stage workflow that mirrors real-world software engineering practices: Issue Understanding, Fault Localization, and Patch Generation. Each stage is enhanced by the MCTS framework~\cite{antoniades2024swesearch} and informed by the experiences retrieved from the experience bank for the current problem.

The MCTS process unfolds as a search through a tree where nodes represent states of the codebase and edges represent actions (\texttt{Search} for code exploration, \texttt{View} for context examination, and \texttt{Edit} for code modification). At each step, the agent selects actions based on a modified Upper Confidence Bound for Trees (UCT) criterion that balances exploiting known high-reward paths with exploring less-visited states. Our experience-enhanced framework augments this standard MCTS exploration by retrieving relevant historical knowledge at critical decision points, providing contextual guidance that informs both action selection and value assessment. When the agent encounters decision nodes during tree expansion, semantically similar experiences from past resolution attempts are dynamically retrieved and integrated into the exploration strategy. This transforms the traditional trial-and-error nature of MCTS into a systematic, knowledge-driven process where each exploration step builds upon accumulated expertise rather than starting from scratch.

\subsubsection{Experience Retrieval}
The framework implements a context-aware retrieval system that seamlessly integrates with the tree search process. Before each decision, the agent retrieves $N$ most relevant experiences from the experience bank based on the vector similarities between the new issue type and attributes with keys in the vector database. 

To adapt the experiences to the current context, a rerank agent then selects \(K\) experiences that are deemed helpful for resolving the current issue. 
During the experience reuse stage, the agent analyzes the similarities and differences between past and present issues, generating contextualized guidance that preserves the essence of successful strategies while adapting to new scenarios. 
For comprehension experiences, the agent compares problem statements to suggest strategic approaches for problem comprehension. The detailed prompt to reuse comprehension experiences is as follows:

\begin{tcolorbox}[promptbox={Reuser – Reuse Comprehension Experience Prompt}]
\begin{lstlisting}[basicstyle=\ttfamily\footnotesize\fontsize{8pt}{8pt}\selectfont,
breaklines=true,lineskip=-0.8pt]
You are a knowledgeable issue resolution assistant. Your task is to [[analyze a current issue]] and [[generalize the received experiences into a new insight]] that is applicable to this issue.

You will be given:
- A `problem_statement` describing the current issue
- A past trajectory with:
  - `issue_description`: The description of the past issue
  - `experience`: Either a `perspective` (how this successful trajectory understood this issue) or `reflections` (insights gained from an unsuccessful trajectory)

Your job is to:
1. [[Compare]] the current `problem_statement` with each past trajectory's `issue_description` and `experience`.
2. [[Adapt]] the old experience to the current issue and produce a new applicable experience.
3. [[Identify]] the most likely entry point in the codebase - based on the problem statement - that is critical to resolving the current issue.

You must output a JSON object with exactly one field:
- `new_experience`: A new experience statement tailored to the current issue, based on the old experience. **The more detailed the better**

Your output must strictly follow the JSON format below:
{
    "new_experience": "<the new experience>"
}
\end{lstlisting}
\end{tcolorbox}
For modification experiences, it considers the code environment and safety patterns to inform repair decisions. The detailed prompt to reuse modification experiences is as follows:
\begin{tcolorbox}[promptbox={Reuser – Reuse Modification Experience Prompt}]
\begin{lstlisting}[
  basicstyle=\ttfamily\footnotesize\fontsize{8pt}{8pt}\selectfont,
  breaklines=true,
  lineskip=-0.8pt
]
You are a strategic assistant helping an agent [[improve its next-step instruction]] in a debugging task.

You are given:
- A `problem_statement`: a natural language description of the current software problem
- A `current_code_exploration_history`: The recent exploration steps taken to understand or debug the current codebase. This may include what has been examined, eliminated, or hypothesized so far.
- An `instruction`: the next step the agent is expected to take
- A list of `experiences`: each offering past insights about how to better approach the corresponding issue.

Your task is to:
1. [[Analyze]] how the current `instruction` relates to the given `issue` and `current_code_exploration_history`
2. [[Identify]] useful, transferable, generalized insights from the past experiences of **modification** type
3. Based on those insights, [[rewrite]] the instruction to make it more robust, strategically informed, and better suited to succeed in this situation

### Important Notes
    - <<Focus only on experience of **modification**>>, and ensure the improved instruction aligns with the original goal but incorporates better reasoning or coverage
    - <<NEVER>> add the content that are not related to solving the current problem
    
Output only the following JSON structure:
{
  "enhanced_instruction": "<A single improved and robust instruction, rewritten based on relevant experience of modification type>"
}
\end{lstlisting}
\end{tcolorbox}

\subsubsection{Agent Role Separation}

When integrating multi-facet experiences into MCTS frameworks, we observe that vanilla MCTS tends to overuse \textit{find} actions and lacks initiative in performing actual code modifications. To address this, we introduce a hierarchical dual-agent architecture that separates high-level planning from low-level execution. The process is managed by an \textit{Instructor} and an \textit{Assistant}: the \textit{Instructor} agent acts as a high-level planner, determining the strategic direction of the next action (search, view, modify, or finish), while the \textit{Assistant} operates at a low level, executing the specific actions based on the information provided by the \textit{Instructor}. 

Such a role separation improves issue resolution by enabling more focused and interpretable repair trajectories. It offers two main advantages: (1) The \textit{Instructor}'s decision-making is streamlined, as it no longer handles irrelevant tools or arguments. (2) Unlike vanilla MCTS, where thought and action generation are coupled, our framework decouples them, providing instruction-level control over tool usage. This allows prior experience, particularly in code modification, to be better leveraged by shaping the \textit{Assistant}'s instructions.

\section{Experimental Setup}

\subsection{Research Questions}
We evaluate \approach by addressing the following research questions: 

\noindent\textbf{RQ1:} How effective is \approach in issue resolution compared to other approaches?

\noindent\textbf{RQ2:} How does each component of \approach contribute to its overall performance?



\noindent \textbf{RQ3:} How does the hyperparameters impact the effectiveness of \approach?

\subsection{Datasets}
We evaluate \approach on the SWE-Bench-Verified dataset~\cite{openai2024introducing}, which contains 500 verified issues from SWE-bench~\cite{jimenez2024swebench}. Software engineering tasks provide a compelling testbed for investigating agent behavior, as they inherently involve complex reasoning, strategic decision-making, and dynamic interaction with the environment (Jimenez et al., 2024). The SWE-bench-verified benchmark exemplifies these challenges by presenting agents with authentic software bugs that require multi-step solutions: understanding natural language issue descriptions, navigating and analyzing the codebase, proposing plausible modifications, and verifying their fixes through test execution.

We adopt the SWE-Bench-Verified in particular because it focuses on issues with human-verified ground truth patches, thereby reducing label noise and ensuring higher evaluation reliability. This allows for more accurate assessment of an agent's true capability to resolve real-world software issues, without confounding effects from noisy or ambiguous labels.

All baseline methods are evaluated on the same dataset.

\subsection{Baselines}
We compare \approach with the following baselines:
\begin{itemize}[leftmargin=0.5cm]
    \item \textbf{Agentless}~\cite{xia2024agentless}: A non-agentic pipeline that decomposes the repair process into distinct phases of localization, repair, and patch validation.
    \item \textbf{SWE-Agent}~\cite{yang2024sweagenta}: A custom agent-computer interface enabling LM agents to interact with repository environments through defined actions.
    \item \textbf{SWE-Search}~\cite{antoniades2024swesearch}: A state-of-the-art repository-level issue resolution agent that uses Monte Carlo Tree Search (MCTS) to explore the solution space.
    \item \textbf{OpenHands}~\cite{wang2024openhands}: An open-source platform for building general-purpose AI agents that solve software and web tasks through code, terminal, and browser interaction.
    \item \textbf{EvoCoder}~\cite{lin2024llms}: A multi-agent continuous learning framework that leverages experience-driven reflection to iteratively refine problem-solving strategies from previously resolved cases.
    \item \textbf{SAGE}~\cite{hayashi2025self}: Another experience-guided framework built on OpenHands, SAGE distills grounded action trajectories into plan abstractions and refines policies through experience.
\end{itemize}


\subsection{Implementation Details}


We implement \approach by extending the SWE-Search~\cite{antoniades2024swesearch} framework with our components. We employ both \texttt{DeepSeek-V3-0324}~\cite{deepseek-ai2025deepseekv3} and \texttt{Claude-4-Sonnet}~\cite{anthropic2025claude4} as our agent models. For \texttt{DeepSeek-V3-0324}, the agent is configured with a temperature of 0.7 and a maximum of 20 iterations, while the remaining configurations follow SWE-Search~\cite{antoniades2024swesearch}. We further set the maximum number of finished nodes to 2, so that the agent stops early once two patches are successfully generated. For \texttt{Claude-4-Sonnet}, we directly follow the SWE-Search~\cite{antoniades2024swesearch} setup.
Due to space limitations, additional hyperparameters and prompts are provided in the replication package\footnote{https://anonymous.4open.science/r/SWE-Exp-6FEA}.

During the experience collection phase, we run SWE-Search on the SWE-Bench-Verified benchmark, collect the resulting trajectories, and extract structured experiences from them. 

During experience retrieval, we first identify the top $N=10$ issues that are most relevant in terms of error type based on cosine similarity. Subsequently, a dedicated reranking agent evaluates these candidates and selects $k=1$ most applicable experience to guide the current resolution process. We compute cosine similarity based on embeddings generated by the Multilingual-E5-Large model\footnote{\url{https://huggingface.co/intfloat/multilingual-e5-large-instruct}}. To prevent data leakage, we exclude all experiences originating from the same code repository and generated after the target instance.

\section{Results}

\subsection{RQ1: Effectiveness}

\begin{table}[h]
    \centering
    \caption{Main effectiveness results on SWE-Bench-Verified dataset.}
    \label{tab:results:main}
        \begin{tabular}{@{}llc@{}}
        \toprule
        \textbf{Method} & \quad \quad \quad \textbf{Model} & \textbf{Pass@1}  \\
        \midrule
        Agentless & \includegraphics[height=1em]{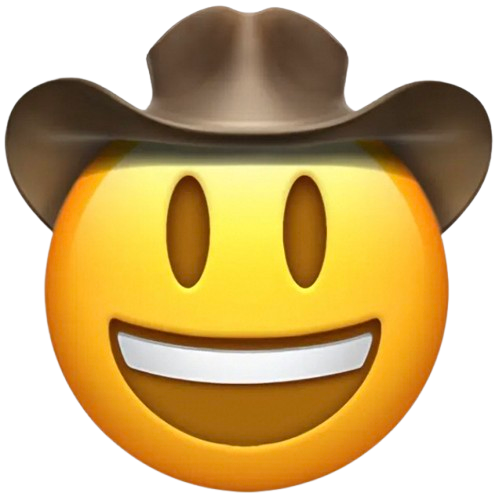} DeepSeek-V3-0324 & 36.6\% \\
        SWE-Agent & \includegraphics[height=1em]{figures/smile.png} DeepSeek-V3-0324 & 38.8\% \\
         & \includegraphics[height=1em]{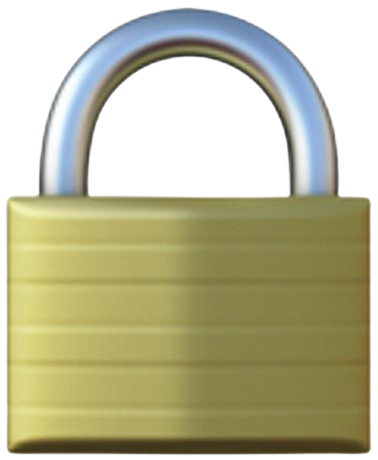} Claude-4 sonnet (20250514) & 66.6\% \\
        SWE-Search & \includegraphics[height=1em]{figures/smile.png} DeepSeek-V3-0324 & 35.4\% \\
                    & \includegraphics[height=1em]{figures/lock.png} Claude-4 sonnet (20250514) & 70.8\% \\
        OpenHands & \includegraphics[height=1em]{figures/smile.png} DeepSeek-V3-0324 & 38.8\% \\ 
                    & \includegraphics[height=1em]{figures/lock.png} Claude-4 sonnet (20250514) & 70.4\% \\
        EvoCoder & \includegraphics[height=1em]{figures/smile.png} DeepSeek-V3-0324 & 38.0\% \\
                & \includegraphics[height=1em]{figures/lock.png} Claude-4 sonnet (20250514) & 67.8\% \\
        SAGE & \includegraphics[height=1em]{figures/lock.png} Claude-4 sonnet (20250514) & 71.6\% \\
        \midrule
        \approach & \includegraphics[height=1em]{figures/smile.png} DeepSeek-V3-0324 & \textbf{42.0\%} \\
                & \includegraphics[height=1em]{figures/lock.png} Claude-4 sonnet (20250514) & \textbf{73.0\%} \\
        \bottomrule
        \end{tabular}
\end{table}

Table~\ref{tab:results:main} presents the comparative performance of \approach against established baselines on the SWE-Bench-Verified dataset. We measure the performance based on widely used metric Pass@1 for issue resolution. This metric captures the proportion of issues that are correctly fixed on the first attempt, in line with the evaluation standards proposed by~\cite{antoniades2024swesearch,yang2024sweagenta}.

Overall, \approach demonstrates strong effectiveness under both open-source and closed-source model settings. Using the open-source \texttt{DeepSeek-V3-0324} backbone, \approach achieves a Pass@1 score of 42.0\%, outperforming all other reported methods under the same model. In particular, it surpasses a diverse set of agent-based frameworks, including SWE-Agent, OpenHands and SWE-Search, which adopt different agentic designs and control strategies.  
We observe similar advantages under the closed-source setting. When evaluated with \texttt{Claude-4 Sonnet}, \approach achieves a Pass@1 score of 73.0\%, consistently outperforming strong agentic baselines such as SWE-Agent, OpenHands, and SWE-Search under the same model. These gains suggest that the benefits of experience-guided orchestration scale up even when the model already has a strong capability.  

Compared to SAGE and EvoCoder, two other experience-driven frameworks that also build upon SWE-Bench-verified trajectories, \approach attains a higher resolution rate under identical model conditions. While SAGE emphasizes stage-wise self-improvement within a single repair process, its experience is primarily localized to the current instance. In contrast, \approach abstracts and reuses experience across trajectories from different repositories, enabling cross-task generalization and more informed orchestration decisions. This distinction allows \approach to better capture recurring repair patterns and failure modes, leading to consistently stronger performance.


The performance improvement stems from effective experience-driven guidance mechanisms. Trajectory-guided problem comprehension experiences enable the Instructor to develop more accurate issue understanding by leveraging patterns from analogous problems, leading to better strategic planning and fault localization hypotheses. Modification-level experiences provide the Assistant with safety patterns and repair strategies that prevent common pitfalls such as incomplete fixes or introducing regressions. This experience-informed approach transforms the repair process from exploratory trial-and-error into systematic, knowledge-guided issue resolution.


\begin{finding}{Finding 1}
    Our proposed method provides significant and consistent performance gains over the baselines, highlighting that experience-guided orchestration enhances performance across both open-source and closed-source models.
\end{finding}

\subsection{RQ2: Ablation Study}

To understand the contribution of each component in \approach, we conduct ablation studies by systematically removing key components. The experiments were conducted within \texttt{DeepSeek-V3-0324} to save API cost~\cite{li2025swe}.

\begin{table}[h]
    \centering
    \caption{Ablation study results.}

    \label{tab:ablation}
    \begin{tabular}{@{}lcc@{}}
    \toprule
    \textbf{Method} & \textbf{Pass@1} & \textbf{$\Delta$} \\
    \midrule
    \textbf{\approach} & \textbf{42.0\%} & - \\
    \quad w/o Comprehension Experience & \textbf{38.8\%} & \textcolor{BrickRed}{\textbf{-3.2\%}} \\
    \quad w/o Modification Experience  &  \textbf{39.4\%} & \textcolor{BrickRed}{\textbf{-2.6\%}} \\
    \quad w/o Dual-Agent  & \textbf{39.8\%}  &  \textcolor{BrickRed}{\textbf{-2.2\%}} \\
    \quad w/o Experiences Extraction  & \textbf{36.0\%}  &  \textcolor{BrickRed}{\textbf{-6.0\%}} \\
    \quad w/o LLM Reranking  & \textbf{38.2\%}  &  \textcolor{BrickRed}{\textbf{-3.8\%}} \\
    \bottomrule
    \end{tabular}
\end{table}

We test the five main components of \approach: 1) \textbf{w/o Comprehension Experience}, which removes comprehension-related experiences and prevents the agent from leveraging prior diagnostic knowledge when analyzing problem statements; 2) \textbf{w/o Modification Experience}, which discards modification-related experiences and thus eliminates guidance on how to safely and robustly apply code changes; 3) \textbf{w/o Dual-Agent}, which collapses the dual-agent architecture into a single agent, removing the separation between strategic reasoning and tactical execution; 4) \textbf{w/o Experiences Extraction}, which disables cross-repository experience abstraction and instead directly uses the raw problem statement and golden patch as in-context demonstrations; and 5) \textbf{w/o LLM Reranking}, which replaces LLM-based experience reranking with simple similarity-based selection.

As shown in Table~\ref{tab:ablation}, the removal of individual components leads to substantial performance drop. The clear impact of removing comprehension experiences (-3.2\%) provides empirical support for our core motivation that existing agents operate as memoryless explorers. These experiences fundamentally transform how agents approach new issues by providing strategic guidance extracted from successful problem-solving patterns observed in our motivating example with CheckboxInput widgets. Without comprehension experiences, agents revert to the problematic behavior as observed in prior analysis—focusing on surface-level symptoms rather than understanding the underlying design patterns. 
Our multi-faceted experience bank design specifically captures these high-level diagnostic insights, enabling the Instructor agent to formulate more accurate hypotheses about root causes from the outset. The smaller but significant impact of modification experiences (-2.6\%) demonstrates their complementary role in our dual-agent architecture, where they guide the Assistant agent in applying proven repair strategies while avoiding common pitfalls such as incomplete fixes or introducing regressions. The dual-agent framework's contribution (-2.2\%) validates our architectural choice to separate strategic reasoning from tactical execution, addressing the mitigating reasoning overload problem that causes vanilla MCTS agents to over-rely on find actions while neglecting actual code modifications.

Removing the experience extraction process—which replaces experience abstraction with directly using the problem statement and golden patch as in-context demonstrations—results in the largest single-component performance drop (-6.0\%). This indicates that abstracting experiences across repositories is crucial for effective knowledge transfer and problem-solving. 
Separately, replacing LLM-based reranking with simple similarity-based selection reduces Pass@1 to 38.2\% (-3.8\%), demonstrating that intelligent selection of relevant experiences is important for achieving high performance.
Together, these results show that while comprehension, modification, and coordination provide the core capabilities, both experience abstraction and intelligent selection play critical, complementary roles in our approach. 

These results confirm that both comprehension and modification experiences contribute positively to system performance, with trajectory-guided problem comprehension playing a slightly more influential role. Even when only one type of experience is used, the system maintains most of its original performance and still achieves improvements over the baseline. Overall, the incorporation of hierarchical experience bank provides consistent and additive gains, validating our design for structured, stage-specific knowledge reuse.

\begin{finding}{Finding 2}
The removal of individual components leads to sustantial performance drop, indicating the importance of the experience-driven issue resolution mechanism.
\end{finding}

\subsection{RQ3: Impact of Hyperparameters}

In this section, we analyze the impact of two hyperparameters, namely, the number of experiences and the size of the experience bank, on the performance of \approach. Here, the number of experiences refers to how many past experiences the agent leverages in attempting to solve each individual instance, while the size of the experience bank denotes the total number of candidate experiences available for retrieval. To save API costs while obtaining reliable results, we conduct experiments on \texttt{DeepSeek-V3-0324}, one of the most popular open-sourced LLMs. 



\begin{wrapfigure}{r}{0.48\textwidth}
  \centering
  \includegraphics[width=0.8\linewidth]{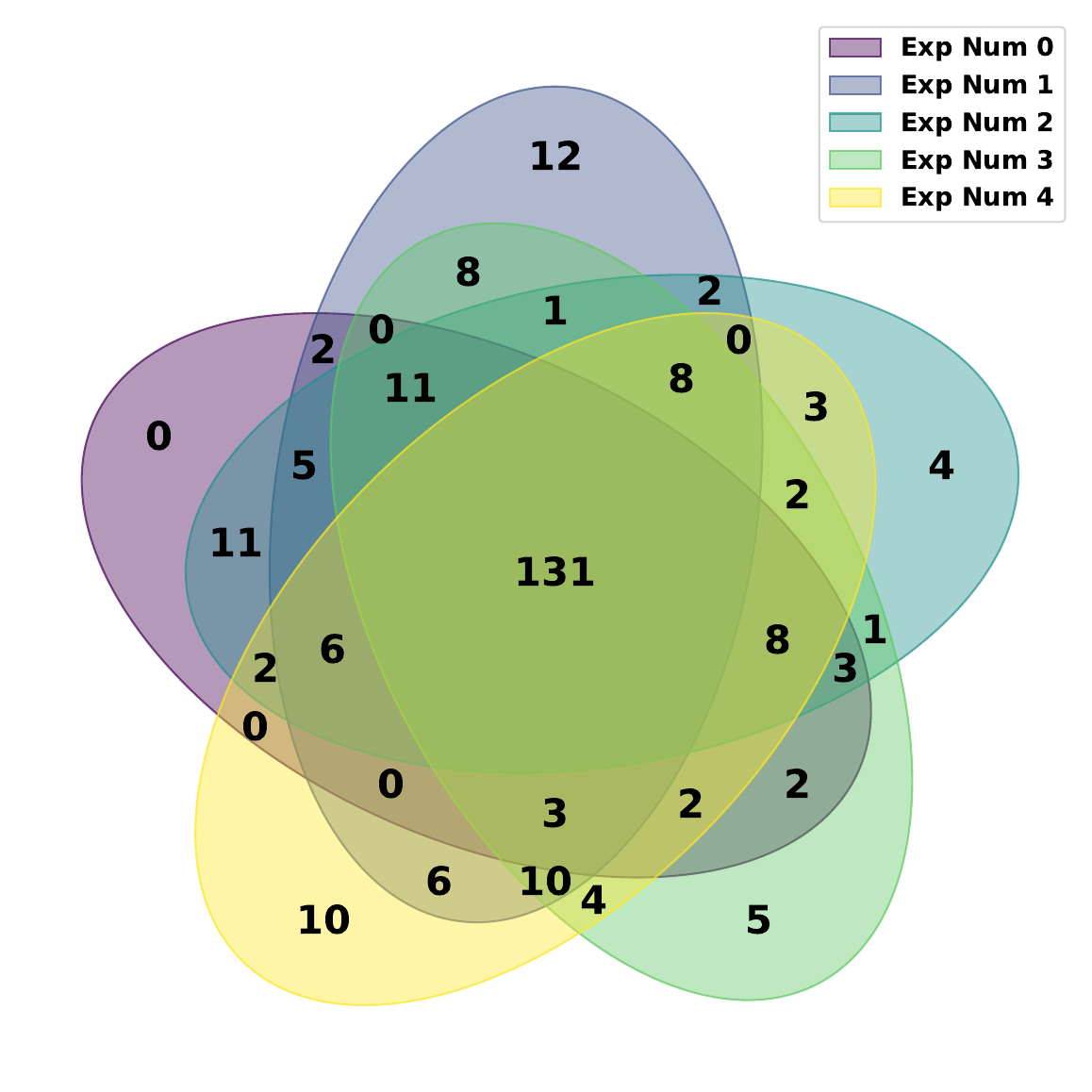}
  \caption{Unique Issues Resolved under Varying Experience Settings.}
  \label{fig:venn}
\end{wrapfigure}

\begin{figure}[h]
    \centering
    \begin{subfigure}[b]{0.48\linewidth}
        \centering
        \includegraphics[width=\linewidth]{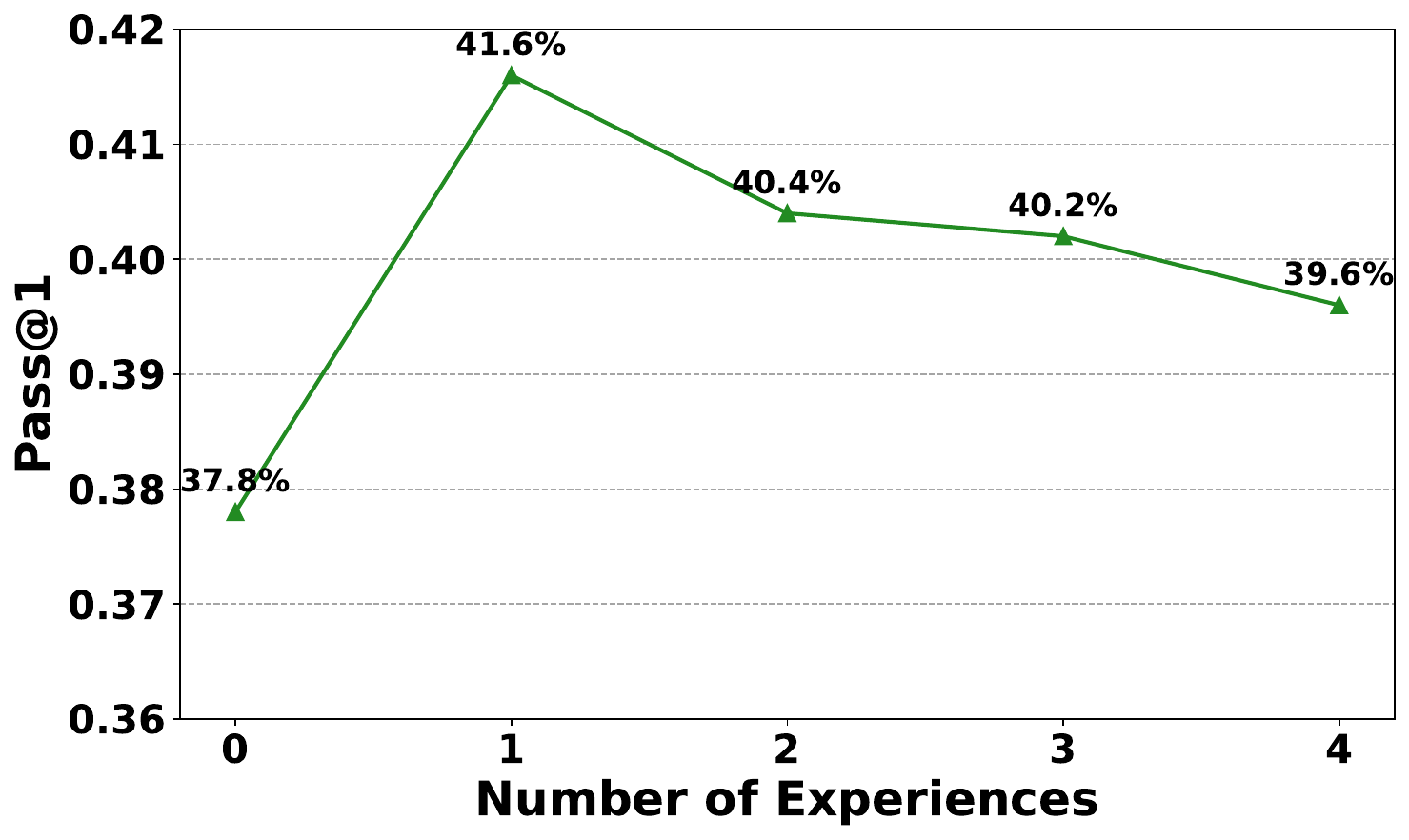}
        \caption{Impact of the number of experiences}
        \label{fig:exp_pass1}
    \end{subfigure}
    \begin{subfigure}[b]{0.48\linewidth}
        \centering
        \includegraphics[width=\linewidth]{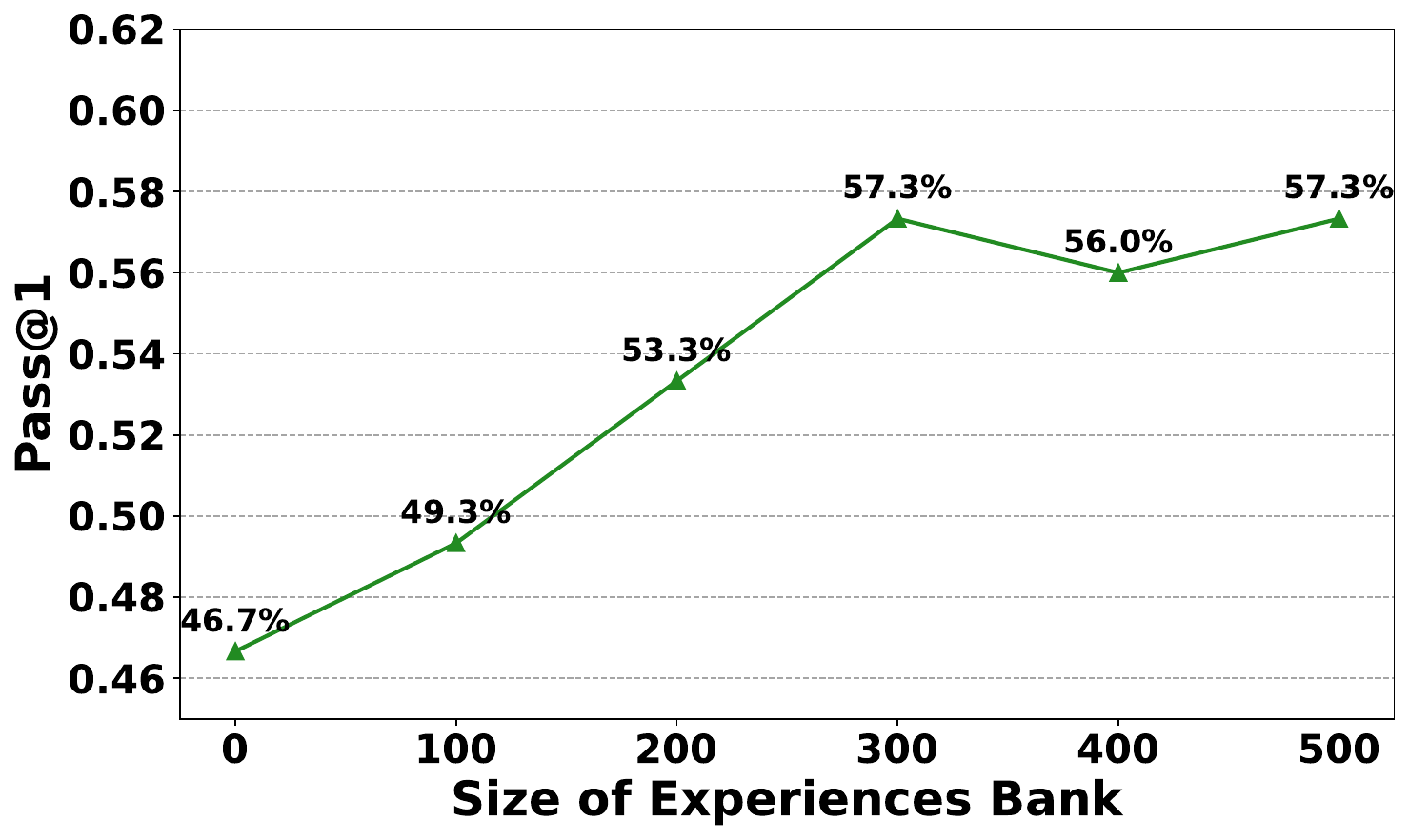}
        \caption{Impact of the size of the experience bank}
        \label{fig:bank_size}
    \end{subfigure}
    \caption{Analysis of Experience Bank Parameters on Issue Resolution Performance}
    \label{fig:exp_analysis}
\end{figure}

We vary the number of experiences from 0 to 4, where 0 means no experience is used. As the results in Figure~\ref{fig:exp_pass1} show, the relationship between experience number and performance demonstrates an increase-then-stable trend. Without any experience, the system achieves 37.8\% Pass@1. The performance peaks at 42.0\% when exactly 1 experience is used, while dropping as the number continues to increase.
This pattern demonstrates that while relevant experiences can significantly enhance performance, excessive guidance can introduce cognitive burden or conflicting information to the agent. This finding underscores the importance of selective experience retrieval and highlights the need for quality over quantity in experience-driven agent systems.

When comparing the resolved instances across different numbers of retrieved experiences (from 0 to 4), we observe substantial variability in the subsets of issues successfully solved. As illustrated in Figure~\ref{fig:venn}, a large number of issues (131) are consistently resolved across all settings. When utilizing past experiences, each experience demonstrates the ability to uniquely resolve specific instances. This suggests that mitigating the misleading influences within individual experiences, while allowing the accumulation of experience to contribute positively, may further enhance the agent's ability to resolve issues.

We further investigate how the size of the experience bank influences the performance of \approach. To reduce computational and API costs while allowing for controlled evaluation of hyperparameter effects, experiments are conducted on a representative subset~\cite{xia2025live, li2025swe, wang2025huxleygodelmachinehumanlevelcoding, zhang2025darwin} 
of 75 instances, with \texttt{DeepSeek-V3-0324}, consisting of 25 issues each from Django, SymPy, and Sphinx. As illustrated in Figure~\ref{fig:bank_size}, we evaluate the impact of experience bank growth by progressively expanding the bank in increments of 100 experiences, following their chronological order of acquisition.

The results show a clear positive trend in performance as the experience bank expands. Specifically, Pass@1 improves steadily as the number of experiences increases, with the most notable gains observed up to approximately 300 experiences. Beyond this point, performance enters a saturation regime, where additional experiences yield diminishing returns, and Pass@1 exhibits only minor fluctuations of 1–2 points. This behavior suggests that once sufficient coverage of common bug patterns and repair strategies is achieved, further accumulation of experiences provides limited incremental benefit.



\begin{finding}{Finding 3}
The hyperparameters have a strong correlation to the performance. Retrieving one single experience is sufficient to achieve optimal performance.
Incorporating a richer set of experiences consistently improves performance.
\end{finding}

\subsection{Case Study}
\begin{figure*}[t]
    \centering
    \includegraphics[width=0.9\textwidth]{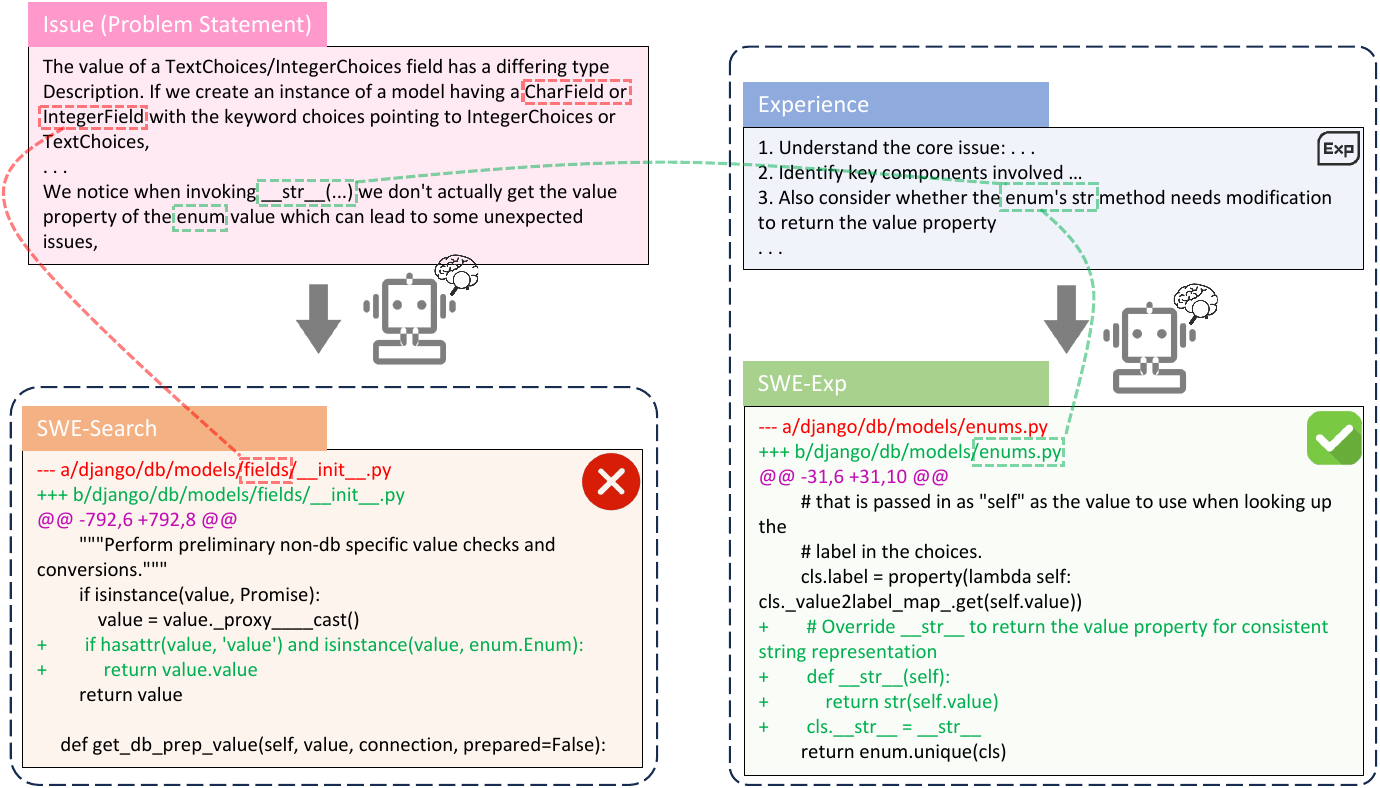}
    \vspace{-0.1cm}
    \caption{Case study of \approach on instance \textit{django-11964}}
    \vspace{-0.2cm}
    \label{fig:case-study}
\end{figure*}

To further verify the effectiveness of \approach in real-world scenarios, We compare two agent trajectories on the same SWE-bench instance—with and without experience reuse—to demonstrate how retrieved experiences influence the agent's decision-making and contribute to successful repair. The results are shown in Figure~\ref{fig:case-study}. 

This case examines a fault related to TextChoices and IntegerChoices, which constitute the core focus of the problem statement. Although the context briefly mentions CharField and IntegerField, the key concern lies in how enum class behaves when converted to string. Specifically, invoking str function on such enum values doesn't yield expected value attribute.

An initial attempt to address this issue by agent specialization involved modifying the get\_db\_prep\_value method within Django's model field handing code. This patch introduced a conditional check to manually extract the value attribute from enum class. Although this solution fixed the immediate problem, it did so in the wrong place. The agent produced a patch that attempts to handle enum conversion within the get\_prep\_value method; however, this modification fails to resolve the actual issue. The root causes of this failure is that: misleading surface-level correlations — the agent incorrectly associated the need for enum conversion with the get\_prep\_value method based on its docstring, without accounting for its actual invocation context and the emphasis in the problem statement.

In contrast, the subsequent fix - developed after incorporating comprehension experience — identified the enum's \_\_str\_\_ method as the appropriate point of intervention. The patch defined the \_\_str\_\_ method within the enum definition to return self.value, thereby ensuring consistent and intuitive string representations throughout the framework. 

This case demonstrates the practical value of transferring knowledge across repositories. Without exposure to prior examples, especially those involving similar symptoms but differing root causes, the model might have repeated the same architectural mistake. However, by leveraging cross-repository experience, it was able to identify the correct point of intervention and propose a solution that was both technically sound and idiomatic to Django's codebase. 

\section{Discussion}

\subsection{Data Leakage}
One critical concern about our experience-driven framework is the threat of knowledge leakage~\cite{zhang2025swebench}. Specifically, in datasets such as SWE-bench, multiple instances from the same repository may correspond to closely related or even identical buggy code segments. If experiences are retrieved from the same repository as the target instance, especially via similarity-based matching, there is a high risk that the agent leverages repository-specific signals or implicitly accesses ground-truth-relevant information. This can lead to artificially inflated performance and fails to demonstrate the true generalizability of the extracted experiences. 
To avoid data leakage, experience retrieval strictly respects both repository boundaries and temporal order: for each instance, we exclude all trajectories from the same repository that are temporally posterior to the current issue. This setup simulates a realistic setting in which agents can only leverage past experiences, and enables a fair assessment of the generalization capability of our experience-driven framework. 
Our experimental results further support the generalizability of our extracted experiences across repositories, as shown in Table~\ref{tab:results:main}. 

\subsection{Quality of Extracted Experience}
We manually verified the correctness of LLM-extracted experiences on a subset of 75 instances, ensuring that the extracted experiences faithfully reflect the underlying trajectories and problem contexts. We then conducted targeted experiments to study different strategies for incorporating these experiences. We also found that, although experiences have been shown to enhance the ability of agent, they could introduce misleading thoughts—particularly at the problem comprehension stage. In early implementations, we allowed the \textit{Instructor} to explicitly cite comprehension experiences as part of its thinking and instructions. While this made the decision process interpretable, it led to too much dependence: the \textit{Instructor} continued using experience even after enough environment exploration, sometimes applying inappropriate strategies. Therefore adding experience as message context—without forcing instructor to use experiences—was the most effective, avoiding inflexible or misleading in the late issue resolution. This illustrates the misleading nature of experience during the comprehension stage: even when the environment has already been sufficiently explored, the agent may continue relying on past experience, leading to inappropriate strategies. This tendency also aligns with a broader trend observed in our quantitative evaluation in Figure~\ref{fig:exp_pass1}: increasing the number of experiences beyond one led to a steady decline in performance. While a single experience boosted Pass@1 from 37.2\% to 42.0\%, adding more examples degraded performance, dropping to 39.8\% with two experiences and further declining to 39.0\% with four experiences. These results suggest that excessive experience may impair the agent's ability to focus and generalize effectively to the current issue.

Moreover, increasing the number of past trajectories used for experience generalization tended to introduce irrelevant or conflicting information, which negatively impacted the agent's effectiveness on the current issue, as the model found it harder to focus on the most relevant information and was more likely to rely on irrelevant or confusing information. In contrast, modification experience showed higher robustness: since the specific direction of the modification instruction is already decided by the Instructor, the Assistant can better assess whether a given experience makes sense or not. 
Overall, providing one single relevant experience alongside the interaction history yields the best performance for the Instructor.

\subsection{Cost Analysis}

Table~\ref{tab:efficiency} compares the efficiency of \approach against SWE-Search under the \texttt{DeepSeek-V3-0324} setting. Despite extending with more modules such as the dual-agent architecture and the experience mechanism, the overall overhead introduced by \approach remains modest. Concretely, the average token consumption increases only slightly (203.3K vs.\ 189.1K), while the average API usage cost remains nearly unchanged (\$0.13 vs.\ \$0.12), indicating that the additional reasoning and coordination steps do not substantially inflate token-level expenses.

In terms of runtime, the experience mechanism introduces an additional overhead of approximately 37 seconds per instance. Nevertheless, the total wall-clock time increases marginally (15min49s vs.\ 12min37s), suggesting that the retrieval component constitutes a relatively small fraction of the end-to-end execution time.

Overall, these results demonstrate that the substantial performance gains of \approach are achieved with minimal additional computational and monetary cost. 


\begin{table}[h]
\centering
\caption{Comparison of Time and Token Costs between SWE-Search and \approach.}
\label{tab:efficiency}
\begin{tabular}{@{}lcc@{}}
\toprule
\textbf{Metrics} & \textbf{SWE-Search} & \textbf{\approach} \\
\midrule
\textbf{Average Token Costs} & 189.1K & 203.3K \\
\textbf{Average API Costs} & \$0.12 & \$0.13 \\
\textbf{Average Wall Time} & 12min37s & 15min49s \\
\textbf{Average Retrieval and Rerank time} & 0s & 37.5s \\
\bottomrule
\end{tabular}
\end{table}

\subsection{Limitations and Future Directions}

While these results are promising, \approach still faces several limitations. Its effectiveness depends on the quality of extracted experiences and their relevance to the target issue; if the retrieved knowledge fails to align with the current problem's semantics, performance may degrade. Moreover, the agent currently lacks a robust mechanism to assess the applicability of prior experiences in novel contexts, which may lead to inappropriate reuse or misleading to irrelevant patterns.

Future work may address these challenges by developing more robust experience extraction methods that better filter noise and identify transferable knowledge patterns. In addition, exploring more accurate retrieval and alignment techniques, incorporating confidence estimation or applicability scoring, and integrating formal verification can further enhance the reliability and adaptability of experience-driven agents in dynamic and unfamiliar code environment.

\section{Threats to Validity}
\textbf{Internal.} The first internal threat comes from our reliance on a single underlying language model for all agent interactions. This choice may introduce model-specific biases and limit the generalizability of our findings across different LLM architectures. To address this concern, we follow prior work in automated program repair~\cite{xia2024agentless, yang2024sweagenta} that demonstrates effectiveness on single-model evaluations, and our dual-agent architecture with experience bank can be readily adapted to other state-of-the-art models.
Another internal threat stems from potential data leakage, where SWE-bench instances may have been included in the training data of our underlying language model. While \texttt{DeepSeek-V3-0324} is open-source, its training data composition is not publicly disclosed, making it impossible to verify potential overlap with our evaluation dataset. In the experiments, our approach shows consistent improvements over strong baselines that use the same underlying models, indicating that gains arise from our architectural innovations rather than training data advantages.

\noindent\textbf{External.} The primary external threat concerns the generalizability of our approach beyond Python repositories and the specific issue types present in SWE-bench. Our evaluation focuses exclusively on Python-based open-source projects, limiting our ability to demonstrate cross-language effectiveness. However, our approach is fundamentally language-agnostic, as it captures high-level issue resolution patterns like problem comprehension and modification experiences rather than language-specific syntax or semantics. The \approach framework should theoretically transfer to other programming languages, making cross-language evaluation a promising direction for future work.

\section{Related Work}

\subsection{Repository-Level Issue Resolution}
Repository-level issue resolution automatically identifies and resolves issues across multiple files within a software project, requiring understanding of complex dependencies and maintaining code consistency~\cite{jimenez2024swebench}. Recent approaches leverage large language models to develop solution frameworks that can be categorized into agentic and non-agentic paradigms. Agentic frameworks treat language models as autonomous agents that step-by-step interact with code environments, with SWE-Agent~\cite{yang2024sweagenta} introducing a foundational agent-computer interface for repository-level interactions. Building on this foundation, several systems have enhanced specific capabilities: AutoCodeRover~\cite{zhang2024autocoderover} and SpecRover~\cite{ruan2024specrover} focus on improved localization and agent support mechanisms, while OpenHands CodeAct~\cite{lv2024codeact} provides comprehensive tooling frameworks. SWE-Search~\cite{antoniades2024swesearch} uses Monte Carlo Tree Search for systematic solution space exploration, 
LocAgent~\cite{chen2025locagent} is a graph-guided LLM agent that performs multi-hop reasoning over a codebase graph to accurately localize relevant code elements for a given issue. Building upon this work, SWE-Debate~\cite{li2025swe} employs multiple reasoning chains combined with a competitive debate mechanism among specialized agents. This design allows SWE-Debate to explore multiple fault propagation traces simultaneously, leading to more accurate issue localization and more informed repair planning. SAGE~\cite{hayashi2025self} enables self‑improving behavior by inducing high‑level plan abstractions from the execution trajectories of the current instance and using these abstractions as contextual guidance to refine subsequent agent policies. 

Non-agentic pipelines focus on specialized execution workflows, with Agentless~\cite{xia2024agentless} decomposing repair into distinct phases of localization, repair, and validation. CodeMonkeys~\cite{ehrlich2025codemonkeys} explores scaling test-time compute through iterative codebase editing with concurrent testing, while recent work~\cite{jiang2025putting} demonstrates that long-context language models with proper prompting can compete with complex agent systems. Training-based approaches have emerged to create SWE-bench-like instances for specialized fine-tuning~\cite{pham2025swesynth, pan2024training, yang2025swesmith}, with MCTS-Refined CoT~\cite{wang2025mctsrefined} using Monte Carlo Tree Search and reflection mechanisms to generate high-quality training data for substantial performance improvements.




Despite remarkable progress in performance results, existing approaches face several critical limitations that hinder their practical effectiveness. Current evaluations rely mainly on static offline datasets, raising concerns about solution memorization and configuration-specific optimizations rather than genuine algorithmic advances~\cite{zhang2025swebench}. While graph-based methods demonstrate effective fault localization and Monte Carlo Tree Search-based exploration shows potential for higher-quality fixes, these methods often provide limited improvements in patch quality due to substantial computational costs and frequent failure to identify correct solutions after extensive search~\cite{jiang2025cosil, antoniades2024swesearch}. Analysis of agent behavior reveals common failure patterns including overthinking and premature disengagement that further limit effectiveness~\cite{cuadron2025danger}. Most critically, existing approaches lack systematic methods to learn from repair experiences, resulting in repeated exploration of failed strategies and missed opportunities to leverage successful patterns from previous attempts~\cite{zhao2024expelc, feng2025get, tang2025agent}. 

EvoCoder~\cite{lin2024llms} introduces a promising multi-agent continuous learning framework for issue code reproduction that uses reflection mechanisms allowing LLMs to continuously learn from previously resolved problems and dynamically improve strategies for new challenges. 
SAGE~\cite{hayashi2025self} proposes a test-time adaptation framework that enables language-model-powered agents to improve their planning and execution over time by inducing concise plan abstractions from their own task experience and conditioning subsequent policy decisions on these abstractions to refine behavior and performance.
Building on this valuable insight of leveraging historical experiences, our \approach further extends the experience-driven paradigm to the complete repository-level issue resolution workflow. \approach introduces a comprehensive experience-enhanced framework that captures and leverages structured experiences across multiple stages of the issue resolution process—from initial problem comprehension to final code modification. Crucially, these experiences are not limited to a single repository; \approach enables the aggregation and reuse of trajectories collected from heterogeneous external codebases, allowing agents to transfer problem-solving knowledge across repositories with differing architectures and domains. Through systematic distillation of multi-faceted experiences (problem comprehension and modification patterns) and their application via a dual-agent architecture, \approach transforms the entire repair workflow from isolated problem-solving into strategic, experience-guided issue resolution.

\subsection{Experience Enhanced AI Agents}

AI agents have fundamentally transformed how we approach complex computational tasks by providing autonomous reasoning and decision-making capabilities that can adapt to diverse problem contexts~\cite{huang2023agentcoder, shi2024code, du2024improving, wu2023autogen, zhang2025kabb, zhang2025gam,wang2025position}. In order to enhance their ability to accumulate and leverage knowledge from past experiences, experience-enhanced agent architectures are proposed~\cite{xie2023olagpt, packer2024memgpt, qian2023experiential, ma2025thinking, ma2025sorft}.

Early foundational work in experience-enhanced AI agents focused on developing human-like memory systems for better long-term interactions~\cite{ zhang2024agent, tang2025agent}. OlaGPT~\cite{xie2023olagpt} introduced cognitive simulation by adding memory and learning from mistakes to copy human-like thought processes. Think-in-Memory (TiM)~\cite{liu2023thinkinmemory} introduced a two-stage framework for recalling thoughts before generation and post-thinking for memory updates. This enables LLMs to maintain evolved memory without repeated reasoning. MemoryBank~\cite{zhong2024memorybank} separated long-term and short-term memory types to create more natural human-machine interactions, while MemGPT~\cite{packer2024memgpt} used hierarchical storage levels with context priority strategies for extended information management. OpenAI's ChatGPT also added memory functionality through external memory layers to store user-specific information across sessions~\cite{openai2024memory}.
These foundational approaches showed the importance of persistent memory systems but mainly focused on conversational contexts rather than task-specific problem-solving.

Modern experience-based learning frameworks have evolved to capture and use procedural knowledge from agent interactions~\cite{wang2024agent, ma2025sorft, zhang2024agent}. ExpeL~\cite{zhao2024expelc} introduced autonomous experience gathering through natural language insights with weighted management systems (ADD, EDIT, UPVOTE, DOWNVOTE) for non-parametric learning. Building on this, AgentRR~\cite{feng2025get} introduced comprehensive record-and-replay systems that capture both environmental interactions and internal decision processes. AutoGuide~\cite{fu2024autoguide} automatically generates context-aware guidelines from offline experiences using contrastive learning techniques. Advanced frameworks like CAIM~\cite{westhausser2025caim} implement advanced cognitive AI-inspired architectures with specialized Memory Controller, Memory Retrieval, and Post-Thinking modules. Recent approaches emphasize learned routine development, with ExACT~\cite{yu2025exact} combining Reflective Monte Carlo Tree Search with vector database storage for dynamic search efficiency improvement. Self-improving coding systems~\cite{robeyns2025selfimproving} achieve autonomous code editing through LLM-driven reflection mechanisms. 
However, existing frameworks mainly target general-purpose tasks and lack domain-specific optimizations for software engineering scenarios~\cite{zhang2024autocoderover, xia2024agentless}. \approach addresses this limitation by developing specialized experience architectures designed for repository-level issue resolution, capturing both strategic repair workflows and detailed code-level patterns.

\section{Conclusion}
We presented \approach, an experience-enhanced framework that transforms repository-level issue resolution from isolated exploration into experience-driven processes. By capturing and distilling knowledge from both successful and failed repair trajectories at multiple levels including comprehension and modification experiences, our dual-agent architecture leverages historical insights to guide strategic planning and tactical execution.
Experimental evaluation on SWE-bench demonstrates significant effectiveness, achieving a Pass@1 score of 73.0\%, establishing a new paradigm where automated agents systematically accumulate and leverage knowledge rather than relying on trial-and-error exploration. Future work can explore more advanced experience extraction mechanisms and integration with formal verification techniques to further enhance automated software engineering capabilities.

\newpage
\section*{Data Availability}
Our code and data are available at \url{https://anonymous.4open.science/r/SWE-Exp-6FEA}.

\bibliographystyle{ACM-Reference-Format}
\bibliography{ref}

@misc{antoniades2024swesearch,
  title = {{{SWE-Search}}: {{Enhancing Software Agents}} with {{Monte Carlo Tree Search}} and {{Iterative Refinement}}},
  shorttitle = {{{SWE-Search}}},
  author = {Antoniades, Antonis and {\"O}rwall, Albert and Zhang, Kexun and Xie, Yuxi and Goyal, Anirudh and Wang, William},
  year = {2024},
  month = dec,
  number = {arXiv:2410.20285},
  eprint = {2410.20285},
  publisher = {arXiv},
  urldate = {2024-12-18},
  archiveprefix = {arXiv}
}

@misc{chen2024coder,
  title = {{{CodeR}}: {{Issue Resolving}} with {{Multi-Agent}} and {{Task Graphs}}},
  shorttitle = {{{CodeR}}},
  author = {Chen, Dong and Lin, Shaoxin and Zeng, Muhan and Zan, Daoguang and Wang, Jian-Gang and Cheshkov, Anton and Sun, Jun and Yu, Hao and Dong, Guoliang and Aliev, Artem and Wang, Jie and Cheng, Xiao and Liang, Guangtai and Ma, Yuchi and Bian, Pan and Xie, Tao and Wang, Qianxiang},
  year = {2024},
  month = jun,
  number = {arXiv:2406.01304},
  eprint = {2406.01304},
  primaryclass = {cs},
  publisher = {arXiv},
  urldate = {2025-07-01},
  archiveprefix = {arXiv}
}

@misc{chen2025locagent,
  title = {{{LocAgent}}: {{Graph-Guided LLM Agents}} for {{Code Localization}}},
  shorttitle = {{{LocAgent}}},
  author = {Chen, Zhaoling and Tang, Xiangru and Deng, Gangda and Wu, Fang and Wu, Jialong and Jiang, Zhiwei and Prasanna, Viktor and Cohan, Arman and Wang, Xingyao},
  year = {2025},
  month = mar,
  number = {arXiv:2503.09089},
  eprint = {2503.09089},
  primaryclass = {cs},
  publisher = {arXiv},
  urldate = {2025-03-19},
  archiveprefix = {arXiv}
}

@misc{chen2025unveilinga,
  title = {Unveiling {{Pitfalls}}: {{Understanding Why AI-driven Code Agents Fail}} at {{GitHub Issue Resolution}}},
  shorttitle = {Unveiling {{Pitfalls}}},
  author = {Chen, Zhi and Ma, Wei and Jiang, Lingxiao},
  year = {2025},
  month = mar,
  number = {arXiv:2503.12374},
  eprint = {2503.12374},
  primaryclass = {cs},
  publisher = {arXiv},
  urldate = {2025-06-10},
  archiveprefix = {arXiv}
}

@misc{cuadron2025danger,
  title = {The {{Danger}} of {{Overthinking}}: {{Examining}} the {{Reasoning-Action Dilemma}} in {{Agentic Tasks}}},
  shorttitle = {The {{Danger}} of {{Overthinking}}},
  author = {Cuadron, Alejandro and Li, Dacheng and Ma, Wenjie and Wang, Xingyao and Wang, Yichuan and Zhuang, Siyuan and Liu, Shu and Schroeder, Luis Gaspar and Xia, Tian and Mao, Huanzhi and Thumiger, Nicholas and Desai, Aditya and Stoica, Ion and Klimovic, Ana and Neubig, Graham and Gonzalez, Joseph E.},
  year = {2025},
  month = feb,
  urldate = {2025-06-10}
}

@misc{deepseek-ai2025deepseekv3,
  title = {{{DeepSeek-V3 Technical Report}}},
  author = {{DeepSeek-AI}},
  year = {2025},
  month = feb,
  number = {arXiv:2412.19437},
  eprint = {2412.19437},
  primaryclass = {cs},
  publisher = {arXiv},
  urldate = {2025-05-16},
  archiveprefix = {arXiv}
}

@inproceedings{du2024improving,
  title = {Improving Factuality and Reasoning in Language Models through Multiagent Debate},
  booktitle = {Proceedings of the 41st {{International Conference}} on {{Machine Learning}}},
  author = {Du, Yilun and Li, Shuang and Torralba, Antonio and Tenenbaum, Joshua B. and Mordatch, Igor},
  year = {2024},
  month = jul,
  series = {{{ICML}}'24},
  volume = {235},
  pages = {11733--11763},
  publisher = {JMLR.org},
  address = {Vienna, Austria},
  urldate = {2025-06-29}
}

@misc{ehrlich2025codemonkeys,
  title = {{{CodeMonkeys}}: {{Scaling Test-Time Compute}} for {{Software Engineering}}},
  shorttitle = {{{CodeMonkeys}}},
  author = {Ehrlich, Ryan and Brown, Bradley and Juravsky, Jordan and Clark, Ronald and R{\'e}, Christopher and Mirhoseini, Azalia},
  year = {2025},
  month = jan,
  urldate = {2025-06-10}
}

@misc{feng2025get,
  title = {Get {{Experience}} from {{Practice}}: {{LLM Agents}} with {{Record}} \& {{Replay}}},
  shorttitle = {Get {{Experience}} from {{Practice}}},
  author = {Feng, Erhu and Zhou, Wenbo and Liu, Zibin and Chen, Le and Dong, Yunpeng and Zhang, Cheng and Zhao, Yisheng and Du, Dong and Hua, Zhichao and Xia, Yubin and Chen, Haibo},
  year = {2025},
  month = may,
  number = {arXiv:2505.17716},
  eprint = {2505.17716},
  primaryclass = {cs},
  publisher = {arXiv},
  urldate = {2025-06-03},
  archiveprefix = {arXiv}
}

@misc{fu2024autoguide,
  title = {{{AutoGuide}}: {{Automated Generation}} and {{Selection}} of {{Context-Aware Guidelines}} for {{Large Language Model Agents}}},
  shorttitle = {{{AutoGuide}}},
  author = {Fu, Yao and Kim, Dong-Ki and Kim, Jaekyeom and Sohn, Sungryull and Logeswaran, Lajanugen and Bae, Kyunghoon and Lee, Honglak},
  year = {2024},
  month = dec,
  number = {arXiv:2403.08978},
  eprint = {2403.08978},
  primaryclass = {cs},
  publisher = {arXiv},
  urldate = {2025-06-03},
  archiveprefix = {arXiv}
}

@misc{huang2023agentcoder,
  title = {{{AgentCoder}}: {{Multi-Agent-based Code Generation}} with {{Iterative Testing}} and {{Optimisation}}},
  shorttitle = {{{AgentCoder}}},
  author = {Huang, Dong and Bu, Qingwen and Zhang, Jie M. and Luck, Michael and Cui, Heming},
  year = {2023},
  month = dec,
  number = {arXiv:2312.13010},
  eprint = {2312.13010},
  publisher = {arXiv},
  urldate = {2023-12-23},
  archiveprefix = {arXiv}
}

@misc{jiang2025cosil,
  title = {{{CoSIL}}: {{Software Issue Localization}} via {{LLM-Driven Code Repository Graph Searching}}},
  shorttitle = {{{CoSIL}}},
  author = {Jiang, Zhonghao and Ren, Xiaoxue and Yan, Meng and Jiang, Wei and Li, Yong and Liu, Zhongxin},
  year = {2025},
  month = mar,
  number = {arXiv:2503.22424},
  eprint = {2503.22424},
  primaryclass = {cs},
  publisher = {arXiv},
  urldate = {2025-04-07},
  archiveprefix = {arXiv}
}

@misc{jiang2025putting,
  title = {Putting {{It All}} into {{Context}}: {{Simplifying Agents}} with {{LCLMs}}},
  shorttitle = {Putting {{It All}} into {{Context}}},
  author = {Jiang, Mingjian and Ruan, Yangjun and Lastras, Luis and Kapanipathi, Pavan and Hashimoto, Tatsunori},
  year = {2025},
  month = may,
  number = {arXiv:2505.08120},
  eprint = {2505.08120},
  primaryclass = {cs},
  publisher = {arXiv},
  urldate = {2025-05-21},
  archiveprefix = {arXiv}
}

@inproceedings{jimenez2024swebench,
  title = {{{SWE-bench}}: {{Can Language Models Resolve Real-world Github Issues}}?},
  shorttitle = {{{SWE-bench}}},
  booktitle = {{{ICLR}}},
  author = {Jimenez, Carlos E. and Yang, John and Wettig, Alexander and Yao, Shunyu and Pei, Kexin and Press, Ofir and Narasimhan, Karthik R.},
  year = {2024},
  month = jan,
  urldate = {2025-06-10}
}

@misc{lin2024llms,
  title = {{{LLMs}} as {{Continuous Learners}}: {{Improving}} the {{Reproduction}} of {{Defective Code}} in {{Software Issues}}},
  shorttitle = {{{LLMs}} as {{Continuous Learners}}},
  author = {Lin, Yalan and Ma, Yingwei and Cao, Rongyu and Li, Binhua and Huang, Fei and Gu, Xiaodong and Li, Yongbin},
  year = {2024},
  month = nov,
  number = {arXiv:2411.13941},
  eprint = {2411.13941},
  primaryclass = {cs},
  publisher = {arXiv},
  urldate = {2025-06-10},
  archiveprefix = {arXiv}
}

@misc{liu2023thinkinmemory,
  title = {Think-in-{{Memory}}: {{Recalling}} and {{Post-thinking Enable LLMs}} with {{Long-Term Memory}}},
  shorttitle = {Think-in-{{Memory}}},
  author = {Liu, Lei and Yang, Xiaoyan and Shen, Yue and Hu, Binbin and Zhang, Zhiqiang and Gu, Jinjie and Zhang, Guannan},
  year = {2023},
  month = nov,
  number = {arXiv:2311.08719},
  eprint = {2311.08719},
  primaryclass = {cs},
  publisher = {arXiv},
  urldate = {2025-06-10},
  archiveprefix = {arXiv}
}

@misc{lv2024codeact,
  title = {{{CodeACT}}: {{Code Adaptive Compute-efficient Tuning Framework}} for {{Code LLMs}}},
  shorttitle = {{{CodeACT}}},
  author = {Lv, Weijie and Xia, Xuan and Huang, Sheng-Jun},
  year = {2024},
  month = aug,
  number = {arXiv:2408.02193},
  eprint = {2408.02193},
  primaryclass = {cs},
  publisher = {arXiv},
  urldate = {2024-12-23},
  archiveprefix = {arXiv}
}

@inproceedings{ma2025improving,
  title = {Improving {{Automated Issue Resolution}} via {{Comprehensive Repository Exploration}}},
  booktitle = {{{ICLR}} 2025 {{Third Workshop}} on {{Deep Learning}} for {{Code}}},
  author = {Ma, Yingwei and Liu, Yue},
  year = {2025},
  month = mar,
  urldate = {2025-06-10}
}

@misc{ma2025sorft,
  title = {{{SoRFT}}: {{Issue Resolving}} with {{Subtask-oriented Reinforced Fine-Tuning}}},
  shorttitle = {{{SoRFT}}},
  author = {Ma, Zexiong and Peng, Chao and Gao, Pengfei and Meng, Xiangxin and Zou, Yanzhen and Xie, Bing},
  year = {2025},
  month = feb,
  number = {arXiv:2502.20127},
  eprint = {2502.20127},
  primaryclass = {cs},
  publisher = {arXiv},
  urldate = {2025-07-08},
  archiveprefix = {arXiv}
}

@misc{ma2025thinking,
  title = {Thinking {{Longer}}, {{Not Larger}}: {{Enhancing Software Engineering Agents}} via {{Scaling Test-Time Compute}}},
  shorttitle = {Thinking {{Longer}}, {{Not Larger}}},
  author = {Ma, Yingwei and Li, Binhua and Dong, Yihong and Jiang, Xue and Cao, Rongyu and Chen, Jue and Huang, Fei and Li, Yongbin},
  year = {2025},
  month = mar,
  number = {arXiv:2503.23803},
  eprint = {2503.23803},
  primaryclass = {cs},
  publisher = {arXiv},
  urldate = {2025-04-07},
  archiveprefix = {arXiv}
}

@misc{openai2024introducing,
  title = {Introducing {{SWE-bench Verified}}},
  author = {{OpenAI}},
  year = {2024},
  urldate = {2025-06-10},
  howpublished = {https://openai.com/index/introducing-swe-bench-verified/}
}

@misc{openai2024memory,
  title = {Memory and New Controls for {{ChatGPT}}},
  author = {{OpenAI}},
  year = {2024},
  month = mar,
  urldate = {2025-06-10},
  howpublished = {https://openai.com/index/memory-and-new-controls-for-chatgpt/}
}

@misc{packer2024memgpt,
  title = {{{MemGPT}}: {{Towards LLMs}} as {{Operating Systems}}},
  shorttitle = {{{MemGPT}}},
  author = {Packer, Charles and Wooders, Sarah and Lin, Kevin and Fang, Vivian and Patil, Shishir G. and Stoica, Ion and Gonzalez, Joseph E.},
  year = {2024},
  month = feb,
  number = {arXiv:2310.08560},
  eprint = {2310.08560},
  primaryclass = {cs},
  publisher = {arXiv},
  urldate = {2025-06-10},
  archiveprefix = {arXiv}
}

@misc{pan2024training,
  title = {Training {{Software Engineering Agents}} and {{Verifiers}} with {{SWE-Gym}}},
  author = {Pan, Jiayi and Wang, Xingyao and Neubig, Graham and Jaitly, Navdeep and Ji, Heng and Suhr, Alane and Zhang, Yizhe},
  year = {2024},
  month = dec,
  urldate = {2025-06-10}
}

@inproceedings{shao2025llms,
  title={Are LLMs Correctly Integrated into Software Systems?},
  author={Shao, Yuchen and Huang, Yuheng and Shen, Jiawei and Ma, Lei and Su, Ting and Wan, Chengcheng},
  booktitle={2025 IEEE/ACM 47th International Conference on Software Engineering (ICSE)},
  pages={1178--1190},
  year={2025},
  organization={IEEE}
}

@misc{pham2025swesynth,
  title = {{{SWE-Synth}}: {{Synthesizing Verifiable Bug-Fix Data}} to {{Enable Large Language Models}} in {{Resolving Real-World Bugs}}},
  shorttitle = {{{SWE-Synth}}},
  author = {Pham, Minh V. T. and Phan, Huy N. and Phan, Hoang N. and Chi, Cuong Le and Nguyen, Tien N. and Bui, Nghi D. Q.},
  year = {2025},
  month = apr,
  number = {arXiv:2504.14757},
  eprint = {2504.14757},
  primaryclass = {cs},
  publisher = {arXiv},
  urldate = {2025-04-24},
  archiveprefix = {arXiv}
}

@misc{robeyns2025selfimproving,
  title = {A {{Self-Improving Coding Agent}}},
  author = {Robeyns, Maxime and Szummer, Martin and Aitchison, Laurence},
  year = {2025},
  month = may,
  number = {arXiv:2504.15228},
  eprint = {2504.15228},
  primaryclass = {cs},
  publisher = {arXiv},
  urldate = {2025-06-03},
  archiveprefix = {arXiv}
}

@misc{ruan2024specrover,
  title = {{{SpecRover}}: {{Code Intent Extraction}} via {{LLMs}}},
  shorttitle = {{{SpecRover}}},
  author = {Ruan, Haifeng and Zhang, Yuntong and Roychoudhury, Abhik},
  year = {2024},
  month = dec,
  number = {arXiv:2408.02232},
  eprint = {2408.02232},
  primaryclass = {cs},
  publisher = {arXiv},
  urldate = {2024-12-23},
  archiveprefix = {arXiv}
}

@misc{shi2024code,
  title = {From {{Code}} to {{Correctness}}: {{Closing}} the {{Last Mile}} of {{Code Generation}} with {{Hierarchical Debugging}}},
  shorttitle = {From {{Code}} to {{Correctness}}},
  author = {Shi, Yuling and Wang, Songsong and Wan, Chengcheng and Gu, Xiaodong},
  year = {2024},
  month = oct,
  number = {arXiv:2410.01215},
  eprint = {2410.01215},
  primaryclass = {cs},
  publisher = {arXiv},
  urldate = {2024-10-05},
  archiveprefix = {arXiv}
}

@inproceedings{shi2024between,
  title={Between Lines of Code: Unraveling the Distinct Patterns of Machine and Human Programmers},
  author={Shi, Yuling and Zhang, Hongyu and Wan, Chengcheng and Gu, Xiaodong},
  booktitle={2025 IEEE/ACM 47th International Conference on Software Engineering (ICSE)},
  pages={51--62},
  year={2024},
  organization={IEEE Computer Society}
}

@misc{wang2024openhands,
  title = {{{OpenHands}}: {{An Open Platform}} for {{AI Software Developers}} as {{Generalist Agents}}},
  shorttitle = {{{OpenHands}}},
  author = {Wang, Xingyao and Li, Boxuan and Song, Yufan and Xu, Frank F. and Tang, Xiangru and Zhuge, Mingchen and Pan, Jiayi and Song, Yueqi and Li, Bowen and Singh, Jaskirat and Tran, Hoang H. and Li, Fuqiang and Ma, Ren and Zheng, Mingzhang and Qian, Bill and Shao, Yanjun and Muennighoff, Niklas and Zhang, Yizhe and Hui, Binyuan and Lin, Junyang and Brennan, Robert and Peng, Hao and Ji, Heng and Neubig, Graham},
  year = {2024},
  month = oct,
  number = {arXiv:2407.16741},
  eprint = {2407.16741},
  publisher = {arXiv},
  urldate = {2024-12-25},
  archiveprefix = {arXiv}
}

@misc{wang2025mctsrefined,
  title = {{{MCTS-Refined CoT}}: {{High-Quality Fine-Tuning Data}} for {{LLM-Based Repository Issue Resolution}}},
  shorttitle = {{{MCTS-Refined CoT}}},
  author = {Wang, Yibo and Peng, Zhihao and Wang, Ying and Wei, Zhao and Yu, Hai and Zhu, Zhiliang},
  year = {2025},
  month = jun,
  number = {arXiv:2506.12728},
  eprint = {2506.12728},
  primaryclass = {cs},
  publisher = {arXiv},
  urldate = {2025-06-29},
  archiveprefix = {arXiv}
}

@misc{wei2025swerl,
  title = {{{SWE-RL}}: {{Advancing LLM Reasoning}} via {{Reinforcement Learning}} on {{Open Software Evolution}}},
  shorttitle = {{{SWE-RL}}},
  author = {Wei, Yuxiang and Duchenne, Olivier and Copet, Jade and Carbonneaux, Quentin and Zhang, Lingming and Fried, Daniel and Synnaeve, Gabriel and Singh, Rishabh and Wang, Sida I.},
  year = {2025},
  month = feb,
  number = {arXiv:2502.18449},
  eprint = {2502.18449},
  primaryclass = {cs},
  publisher = {arXiv},
  urldate = {2025-02-26},
  archiveprefix = {arXiv}
}

@misc{westhausser2025caim,
  title = {{{CAIM}}: {{Development}} and {{Evaluation}} of a {{Cognitive AI Memory Framework}} for {{Long-Term Interaction}} with {{Intelligent Agents}}},
  shorttitle = {{{CAIM}}},
  author = {Westh{\"a}u{\ss}er, Rebecca and Berenz, Frederik and Minker, Wolfgang and Zepf, Sebastian},
  year = {2025},
  month = may,
  number = {arXiv:2505.13044},
  eprint = {2505.13044},
  primaryclass = {cs},
  publisher = {arXiv},
  urldate = {2025-06-03},
  archiveprefix = {arXiv}
}

@misc{wu2023autogen,
  title = {{{AutoGen}}: {{Enabling Next-Gen LLM Applications}} via {{Multi-Agent Conversation}}},
  shorttitle = {{{AutoGen}}},
  author = {Wu, Qingyun and Bansal, Gagan and Zhang, Jieyu and Wu, Yiran and Li, Beibin and Zhu, Erkang and Jiang, Li and Zhang, Xiaoyun and Zhang, Shaokun and Liu, Jiale and Awadallah, Ahmed Hassan and White, Ryen W. and Burger, Doug and Wang, Chi},
  year = {2023},
  month = oct,
  number = {arXiv:2308.08155},
  eprint = {2308.08155},
  primaryclass = {cs},
  publisher = {arXiv},
  urldate = {2023-12-06},
  archiveprefix = {arXiv}
}

@misc{xia2024agentless,
  title = {Agentless: {{Demystifying LLM-based Software Engineering Agents}}},
  shorttitle = {Agentless},
  author = {Xia, Chunqiu Steven and Deng, Yinlin and Dunn, Soren and Zhang, Lingming},
  year = {2024},
  month = jul,
  number = {arXiv:2407.01489},
  eprint = {2407.01489},
  publisher = {arXiv},
  urldate = {2024-07-08},
  archiveprefix = {arXiv}
}

@misc{xie2023olagpt,
  title = {{{OlaGPT}}: {{Empowering LLMs With Human-like Problem-Solving Abilities}}},
  shorttitle = {{{OlaGPT}}},
  author = {Xie, Yuanzhen and Xie, Tao and Lin, Mingxiong and Wei, WenTao and Li, Chenglin and Kong, Beibei and Chen, Lei and Zhuo, Chengxiang and Hu, Bo and Li, Zang},
  year = {2023},
  month = may,
  number = {arXiv:2305.16334},
  eprint = {2305.16334},
  primaryclass = {cs},
  publisher = {arXiv},
  urldate = {2025-06-10},
  archiveprefix = {arXiv}
}

@inproceedings{yang2024sweagenta,
  title = {{{SWE-agent}}: {{Agent-Computer Interfaces Enable Automated Software Engineering}}},
  shorttitle = {{{SWE-agent}}},
  booktitle = {The {{Thirty-eighth Annual Conference}} on {{Neural Information Processing Systems}}},
  author = {Yang, John and Jimenez, Carlos E. and Wettig, Alexander and Lieret, Kilian and Yao, Shunyu and Narasimhan, Karthik R. and Press, Ofir},
  year = {2024},
  month = nov,
  urldate = {2025-06-10}
}

@misc{yang2025swesmith,
  title = {{{SWE-smith}}: {{Scaling Data}} for {{Software Engineering Agents}}},
  shorttitle = {{{SWE-smith}}},
  author = {Yang, John and Leret, Kilian and Jimenez, Carlos E. and Wettig, Alexander and Khandpur, Kabir and Zhang, Yanzhe and Hui, Binyuan and Press, Ofir and Schmidt, Ludwig and Yang, Diyi},
  year = {2025},
  month = apr,
  urldate = {2025-06-10}
}

@misc{yu2025exact,
  title = {{{ExACT}}: {{Teaching AI Agents}} to {{Explore}} with {{Reflective-MCTS}} and {{Exploratory Learning}}},
  shorttitle = {{{ExACT}}},
  author = {Yu, Xiao and Peng, Baolin and Vajipey, Vineeth and Cheng, Hao and Galley, Michel and Gao, Jianfeng and Yu, Zhou},
  year = {2025},
  month = feb,
  number = {arXiv:2410.02052},
  eprint = {2410.02052},
  primaryclass = {cs},
  publisher = {arXiv},
  urldate = {2025-06-03},
  archiveprefix = {arXiv}
}

@inproceedings{zhang2024autocoderover,
  title = {{{AutoCodeRover}}: {{Autonomous Program Improvement}}},
  shorttitle = {{{AutoCodeRover}}},
  booktitle = {Proceedings of the 33rd {{ACM SIGSOFT International Symposium}} on {{Software Testing}} and {{Analysis}}},
  author = {Zhang, Yuntong and Ruan, Haifeng and Fan, Zhiyu and Roychoudhury, Abhik},
  year = {2024},
  month = sep,
  series = {{{ISSTA}} 2024},
  pages = {1592--1604},
  publisher = {Association for Computing Machinery},
  address = {New York, NY, USA},
  urldate = {2024-10-10},
  isbn = {979-8-4007-0612-7}
}

@misc{zhang2025swebench,
  title = {{{SWE-bench Goes Live}}!},
  author = {Zhang, Linghao and He, Shilin and Zhang, Chaoyun and Kang, Yu and Li, Bowen and Xie, Chengxing and Wang, Junhao and Wang, Maoquan and Huang, Yufan and Fu, Shengyu and Nallipogu, Elsie and Lin, Qingwei and Dang, Yingnong and Rajmohan, Saravan and Zhang, Dongmei},
  year = {2025},
  month = may,
  number = {arXiv:2505.23419},
  eprint = {2505.23419},
  primaryclass = {cs},
  publisher = {arXiv},
  urldate = {2025-06-02},
  archiveprefix = {arXiv}
}

@article{zhao2024expelc,
  title = {{{ExpeL}}: {{LLM Agents Are Experiential Learners}}},
  shorttitle = {{{ExpeL}}},
  author = {Zhao, Andrew and Huang, Daniel and Xu, Quentin and Lin, Matthieu and Liu, Yong-Jin and Huang, Gao},
  year = {2024},
  month = mar,
  journal = {Proceedings of the AAAI Conference on Artificial Intelligence},
  volume = {38},
  number = {17},
  pages = {19632--19642},
  urldate = {2025-06-10},
  copyright = {Copyright (c) 2024 Association for the Advancement of Artificial Intelligence}
}

@inproceedings{zhong2024memorybank,
  title = {{{MemoryBank}}: Enhancing Large Language Models with Long-Term Memory},
  shorttitle = {{{MemoryBank}}},
  booktitle = {Proceedings of the {{Thirty-Eighth AAAI Conference}} on {{Artificial Intelligence}} and {{Thirty-Sixth Conference}} on {{Innovative Applications}} of {{Artificial Intelligence}} and {{Fourteenth Symposium}} on {{Educational Advances}} in {{Artificial Intelligence}}},
  author = {Zhong, Wanjun and Guo, Lianghong and Gao, Qiqi and Ye, He and Wang, Yanlin},
  year = {2024},
  month = feb,
  series = {{{AAAI}}'24/{{IAAI}}'24/{{EAAI}}'24},
  volume = {38},
  pages = {19724--19731},
  publisher = {AAAI Press},
  urldate = {2025-06-10},
  isbn = {978-1-57735-887-9}
}

@article{wang2026swe,
  title={SWE-Pruner: Self-Adaptive Context Pruning for Coding Agents},
  author={Wang, Yuhang and Shi, Yuling and Yang, Mo and Zhang, Rongrui and He, Shilin and Lian, Heng and Chen, Yuting and Ye, Siyu and Cai, Kai and Gu, Xiaodong},
  journal={arXiv preprint arXiv:2601.16746},
  year={2026}
}

@inproceedings{wang2025position,
  title={Position bias mitigates position bias: Mitigate position bias through inter-position knowledge distillation},
  author={Wang, Yifei and Xiong, Feng and Wang, Yong and Li, Linjing and Chu, Xiangxiang and Zeng, Daniel Dajun},
  booktitle={Proceedings of the 2025 Conference on Empirical Methods in Natural Language Processing},
  pages={1495--1512},
  year={2025}
}

@article{qian2023experiential,
  title={Experiential co-learning of software-developing agents},
  author={Qian, Chen and Dang, Yufan and Li, Jiahao and Liu, Wei and Xie, Zihao and Wang, Yifei and Chen, Weize and Yang, Cheng and Cong, Xin and Che, Xiaoyin and others},
  journal={arXiv preprint arXiv:2312.17025},
  year={2023}
}

@article{zhang2024agent,
  title={Agent-pro: Learning to evolve via policy-level reflection and optimization},
  author={Zhang, Wenqi and Tang, Ke and Wu, Hai and Wang, Mengna and Shen, Yongliang and Hou, Guiyang and Tan, Zeqi and Li, Peng and Zhuang, Yueting and Lu, Weiming},
  journal={arXiv preprint arXiv:2402.17574},
  year={2024}
}

@inproceedings{zhang2025kabb,
title={{KABB}: Knowledge-Aware Bayesian Bandits for Dynamic Expert Coordination in Multi-Agent Systems},
author={Jusheng Zhang and Zimeng Huang and Yijia Fan and Ningyuan Liu and Mingyan Li and Zhuojie Yang and Jiawei Yao and Jian Wang and Keze Wang},
booktitle={Forty-second International Conference on Machine Learning},
year={2025},
url={https://openreview.net/forum?id=AKvy9a4jho}
}

@article{zhang2025gam,
  title={GAM-Agent: Game-Theoretic and Uncertainty-Aware Collaboration for Complex Visual Reasoning},
  author={Zhang, Jusheng and Fan, Yijia and Lin, Wenjun and Chen, Ruiqi and Jiang, Haoyi and Chai, Wenhao and Wang, Jian and Wang, Keze},
  journal={arXiv preprint arXiv:2505.23399},
  year={2025}
}

@article{tang2025agent,
  title={Agent KB: Leveraging Cross-Domain Experience for Agentic Problem Solving},
  author={Tang, Xiangru and Qin, Tianrui and Peng, Tianhao and Zhou, Ziyang and Shao, Daniel and Du, Tingting and Wei, Xinming and Xia, Peng and Wu, Fang and Zhu, He and others},
  journal={arXiv preprint arXiv:2507.06229},
  year={2025}
}

@article{wang2024agent,
  title={Agent workflow memory},
  author={Wang, Zora Zhiruo and Mao, Jiayuan and Fried, Daniel and Neubig, Graham},
  journal={arXiv preprint arXiv:2409.07429},
  year={2024}
}

@article{hayashi2025self,
  title={Self-Abstraction from Grounded Experience for Plan-Guided Policy Refinement},
  author={Hayashi, Hiroaki and Pang, Bo and Zhao, Wenting and Liu, Ye and Gokul, Akash and Bansal, Srijan and Xiong, Caiming and Yavuz, Semih and Zhou, Yingbo},
  journal={arXiv preprint arXiv:2511.05931},
  year={2025}
}

@online{anthropic2025claude4,
  author  = {Anthropic},
  title   = {Introducing Claude 4},
  date    = {2025-05-22},
  url     = {https://www.anthropic.com/news/claude-4},
  urldate = {2025-12-24}
}

@article{li2025swe,
  title={Swe-debate: Competitive multi-agent debate for software issue resolution},
  author={Li, Han and Shi, Yuling and Lin, Shaoxin and Gu, Xiaodong and Lian, Heng and Wang, Xin and Jia, Yantao and Huang, Tao and Wang, Qianxiang},
  journal={arXiv preprint arXiv:2507.23348},
  year={2025}
}

@article{xia2025live,
  title={Live-SWE-agent: Can Software Engineering Agents Self-Evolve on the Fly?},
  author={Xia, Chunqiu Steven and Wang, Zhe and Yang, Yan and Wei, Yuxiang and Zhang, Lingming},
  journal={arXiv preprint arXiv:2511.13646},
  year={2025}
}

@article{zhang2025darwin,
  title={Darwin Godel Machine: Open-Ended Evolution of Self-Improving Agents},
  author={Zhang, Jenny and Hu, Shengran and Lu, Cong and Lange, Robert and Clune, Jeff},
  journal={arXiv preprint arXiv:2505.22954},
  year={2025}
}

@misc{wang2025huxleygodelmachinehumanlevelcoding,
      title={Huxley-G\"odel Machine: Human-Level Coding Agent Development by an Approximation of the Optimal Self-Improving Machine}, 
      author={Wenyi Wang and Piotr Piękos and Li Nanbo and Firas Laakom and Yimeng Chen and Mateusz Ostaszewski and Mingchen Zhuge and Jürgen Schmidhuber},
      year={2025},
      eprint={2510.21614},
      archivePrefix={arXiv},
      primaryClass={cs.AI},
      url={https://arxiv.org/abs/2510.21614}, 
}

@article{shi2025longcodezip,
  title={LongCodeZip: Compress Long Context for Code Language Models},
  author={Shi, Yuling and Qian, Yichun and Zhang, Hongyu and Shen, Beijun and Gu, Xiaodong},
  journal={arXiv preprint arXiv:2510.00446},
  year={2025}
}

@article{peng2025swe,
  title={SWE-QA: Can Language Models Answer Repository-level Code Questions?},
  author={Peng, Weihan and Shi, Yuling and Wang, Yuhang and Zhang, Xinyun and Shen, Beijun and Gu, Xiaodong},
  journal={arXiv preprint arXiv:2509.14635},
  year={2025}
}
\clearpage
\appendix
\section{Hyperparameters of MCTS}
\label{appendix:mcts}

The Monte Carlo Tree Search (MCTS) algorithm~\cite{jimenez2024swebench} used in this study employs hyperparameters in Table~\ref{tab:mcts-hyperparams}.

\begin{table}[h]
\centering
\caption{MCTS Hyperparameters}
\label{tab:mcts-hyperparams}
\resizebox{\columnwidth}{!}{
\begin{tabular}{@{}llc@{}}
\toprule
\textbf{Hyperparameter} & \textbf{Description} & \textbf{Default} \\
\midrule
\multicolumn{3}{@{}l}{\textit{Main Search Parameters}} \\
\quad c\_param & UCT exploration parameter & 1.41 \\
\quad max\_expansions & Max children per node & 3 \\
\quad max\_iterations & Max MCTS iterations & 20 \\
\quad provide\_feedback & Enable feedback & True \\
\quad best\_first & Use best-first strategy & True \\
\quad value\_function\_temperature & Value function temperature & 0.2 \\
\quad max\_depth & Max tree depth & 20 \\
\addlinespace
\multicolumn{3}{@{}l}{\textit{UCT Score Calculation Parameters}} \\
\quad exploration\_weight & UCT exploration weight & 1.0 \\
\quad depth\_weight & Depth penalty weight & 0.8 \\
\quad depth\_bonus\_factor & Depth bonus factor & 200 \\
\quad high\_value\_threshold & High-value node threshold & 55 \\
\quad low\_value\_threshold & Low-value node threshold & 50 \\
\quad very\_high\_value\_threshold & Very high-value threshold & 75 \\
\quad high\_value\_leaf\_bonus\_constant & High-value leaf bonus & 20 \\
\quad high\_value\_bad\_children\_bonus\_constant & High-value bad children bonus & 20 \\
\quad high\_value\_child\_penalty\_constant & High-value child penalty & 5 \\
\addlinespace
\multicolumn{3}{@{}l}{\textit{Action Model Parameters}} \\
\quad action\_model\_temperature & Action model temperature & 0.7 \\
\addlinespace
\multicolumn{3}{@{}l}{\textit{Discriminator Parameters}} \\
\quad number\_of\_agents & Number of Discriminator Agents & 5 \\
\quad number\_of\_round & Number of debate rounds & 3 \\
\quad discriminator\_temperature & Discriminator temperature & 1 \\
\bottomrule
\end{tabular}
}
\end{table}

\section{Prompt Templates}
In the following section, we provide a comprehensive enumeration of all prompts employed throughout our workflow, including the system prompts used by the dual-agent architecture, the prompts designed for extracting successful and failed experiences, and those used for reusing past experiences. This detailed documentation aims to ensure reproducibility and to highlight the role of prompt engineering in the effectiveness of our method.
\subsection{Instructor}

\begin{tcolorbox}[promptbox={Prompt 1: Instructor Prompt}]
\begin{lstlisting}[basicstyle=\ttfamily\footnotesize, breaklines=true]
You are an autonomous AI instructor with deep analytical capabilities. Operating independently, you cannot communicate with the user but must analyze the past history of interactions with the code repository to generate the next instruction that guides the assistant toward completing the task.

# Workflow to guide assistants in modifying code

Follow these structured steps to understand the task and instruct the assistant to locate context, and perform code modifications.

### 1. Understand the Task
    - Carefully read the <task> to determine exactly what is known and what still needs to be clarified according to the interaction history.
    - Focus on the cause of the <task> and suggested changes to the <task> that have been explicitly stated in the <task>.
    - Compare <task> with the code from the interaction history, determine what additional context (files, functions, dependencies) may be required. Request more information if needed.

### 2. Locate Code
    - Using your analysis, generate instructions to guide assistant to locate the exact code regions to understand or modify.
    - Once the location of the code that needs to be modified is determined, instruct assistant to modify it and provide the exact location.
    - Narrow down the scope of the code you need to look at step by step.

### 3. Modify Code
    - The generated instruction should only focus on the changes needed to satisfy the task. Do not modify unrelated code.
    - The instructions for modifying the code need to refer to the task and the relevant code retrieved, rather than being based on your own guesses.
    - Keep the edits minimal, correct, and localized.
    - If the change involves multiple locations, apply atomic modifications sequentially.

### 4. Iterate as Needed
    - If the task has already been resolved by the existing code modifications, finish the process without making additional changes.
    - If the task is not fully resolved, analyze what remains and focus only on the unresolved parts.
    - Avoid making unnecessary changes to previously correct code modifications. Subsequent edits should strictly target the remaining issues.
    - When modifying the input parameters or return values of a function or class, make sure to update all relevant code snippets that invoke them accordingly.
    - But do not take test into account, just focus on how to resolve the task.
    - Repeat until the task are resolved.

### 5. Complete Task
    - Once the implementation satisfies all task constraints and maintains system integrity:
      - Do not add additional test cases.
      - Stop the task.

# Additional Notes

 * **Think Step by Step**
   - Always document your reasoning and thought process in the Thought section.
   - Only one kind of instruction is generated each step.

 * **Efficient Operation**
   - Use previous observations to inform your next actions.
   - Avoid instructing assistant to execute similar actions as before.
   - Focus on minimal viable steps: Prioritize actions that maximize progress with minimal code exploration or modification.

 * **Never Guess**
   - Do not guess line numbers or code content.
   - All code environment information must come from the real environment feedback.

# Instructor Output Format
For each input, you must output a JSON object with exactly three fields:
    1. thoughts: A natural language description that summarizes the current code environment, previous steps taken, and relevant contextual reasoning.
    2. instructions:
        - One specific and actionable objective for the assistant to complete next. This should be phrased as a goal rather than an implementation detail, guiding what should be achieved based on the current context.
        - Instruction related to modifying the code must strictly refer to the task at the beginning, and you shouldn't guess how to modify.
        - Do not include any instructions related to test cases.
        - The more detailed the better.
    3. context:
        - If the next step involves retrieving additional context according to the previous observations, ensure the context includes the following specific details from the code environment (as applicable):
            -- Exact file path or vague file pattern(e.g., **/dictionary/*.py)
            -- Exact Class names from environment feedback
            -- Exact Function names from environment feedback
            -- Exact Code block identifiers from environment feedback (e.g., method headers, class declarations)
            -- Exact Corresponding line ranges from environment feedback (start_line and end_line)
            -- The span ids of the code you hope to view
        - If the code environment is uncertain or specific classes and functions cannot be retrieved multiple times, 
            -- Only output a natural language query describing the functionality of the code that needs to be retrieved, without exact file, class, function, or code snippets.
        - If the next step needs to modify the code, the context must contain specific file path.
        - If the task is complete, this could return `None`.
        - Don't guess the context, the context must come from the interaction with the code environment.
    4. type: A string indicating the kind of next action required. Must be one of:
        - "search": when more information is needed,
        - "view": when additional context not returned by searches, or specific line ranges you discovered from search results
        - "modify": when you have identified the specific code to be modified or generated from the code environment feedback.
        - "finish": when the task has been solved.

The instructor's output must follow a structured JSON format:
{
  "thoughts": "<analysis and summary of the current code environment and interaction history>",
  "instructions": "<next objective for the assistant and some insights from the previous actions>",
  "context": "<the description or query that summarizes the code environment that needs to be known in the next step>",
  "type": "<search | view | modify | finish>"
}
\end{lstlisting}
\end{tcolorbox}

\subsection{Assistant}

\begin{tcolorbox}[promptbox={Prompt 2: Assistant Prompt}]
\begin{lstlisting}[basicstyle=\ttfamily\footnotesize, breaklines=true]
# Guidelines for Executing Actions Based on Instructions:

1. Analysis First:
    - Read the problem statement in <task> to understand the global goal.
    - Read the instructor’s instruction in <instruction> to understand the next action.
    
2. Analyze Environment, Interaction History and Code Snippet:
    - If the next action requires retrieving more context, carefully extract precise targets from the <environment>. These may include relevant file names, class names, function names, code block identifiers, or corresponding line ranges, depending on what is available in the context.
    - Actions and their arguments from the past interactions are recorded in <history>. Your next action should retrieve content that is not redundant with those previous actions.
    - If the next action involves modifying code, use the <environment> to get the target path and identify the exact code snippet that needs to be changed in <code>, along with its surrounding logic and dependencies. This ensures the modification is accurate, consistent, and context-aware.

2. EVERY response must follow EXACTLY this format:
   Thought: Your reasoning and analysis
   Action: ONE specific action to take

3. Your Thought section MUST include:
   - What you learned from previous Observations
   - Why you're choosing this specific action
   - What you expect to learn/achieve
   - Any risks to watch for

# Action Description
1. **Locate Code**
  * **Primary Method - Search Functions:** Use these to find relevant code:
      * FindClass - Search for class definitions by class name
      * FindFunction - Search for function definitions by function name
      * FindCodeSnippet - Search for specific code patterns or text
      * SemanticSearch - Search code by semantic meaning and natural language description
  * **Secondary Method - ViewCode:** Only use when you need to see:
      * Additional context not returned by searches but in the same file
      * Specific line ranges you discovered from search results
      * Code referenced in error messages or test failures
  
2. **Modify Code**
  * **Fix Task:** Make necessary code changes to resolve the task requirements
  * **Primary Method - StringReplace:** Use this to apply code modifications
    - Replace exact text strings in files with new content
    - The old_str argument cannot be empty.
  * **Secondary Method - CreateFile:** Only use when you need to need to implement new functionality:
    - Create new files with specified content

3. **Complete Task**
  * Use Finish when confident all applied patch are correct and complete.

# Important Guidelines

 * **Focus on the Specific Instruction**
   - Implement requirements exactly as specified, without additional changes.
   - Do not modify code unrelated to the task.

 * **Code Context and Changes**
   - Limit code changes to files in the code you can see.
   - If you need to examine more code, use ViewCode to see it.

 * **Task Completion**
   - Finish the task only when the task is fully resolved.
   - Do not suggest code reviews or additional changes beyond the scope.

# Additional Notes

 * **Think Step by Step**
   - Always document your reasoning and thought process in the Thought section.
   - Build upon previous steps without unnecessary repetition.

 * **Never Guess**
   - Do not guess line numbers or code content. Use ViewCode to examine code when needed.
\end{lstlisting}
\end{tcolorbox}

\subsection{Issue Agent}

\begin{tcolorbox}[promptbox={Prompt 3: Issue Agent Prompt}]
\begin{lstlisting}[basicstyle=\ttfamily\footnotesize, breaklines=true]
You are an expert error classification assistant. Your task is to analyze string-formatted issue reports and identify the type of error they contain.
For each input, you must output a JSON object with exactly two fields:

1. `issue_type`: The generalized error category in the format "<generalized_descriptive_name>Error" (e.g., "SyntaxError", "NullReferenceError")
2. `description`: A brief description (1-2 sentences) of the characteristics of the identified error category

Your output should strictly follow JSON format with the following structure:
{
    "issue_type": "<generalized_descriptive_name>Error", 
    "description": "<the brief description>",
}
\end{lstlisting}
\end{tcolorbox}

\subsection{Issue Comprehension ExpAgent}

\subsubsection{Successful Experience Extraction Prompt}

\begin{tcolorbox}[promptbox={Prompt 4: Issue Comprehension ExpAgent (Success)}]
\begin{lstlisting}[basicstyle=\ttfamily\footnotesize, breaklines=true]
You are a bug resolution expert. You will be given a software issue, the corresponding golden patch and a trajectory that represents how an agent successfully resolved this issue.

## Guidelines
You need to extract two key aspects from this successful trajectory:
1. **perspective** - how this trajectory thought about this issue - that is, how the problem was understood in a way that **led to its successful resolution**. This should be abstract and not name specific code entities.

## Important Notes:
    - Perspective should be at the level of thinking, not specific implementation details.
    - Perspective and reasoning should be expressed in as generalized and abstract terms as possible.
    - Do not include specific object names in perspective.

Your output must strictly follow the JSON format shown below:
{
    "perspective": "<1-2 sentences to describe how this trajectory understood this issue>",
}
\end{lstlisting}
\end{tcolorbox}

\subsubsection{Failed Experience Extraction Prompt}

\begin{tcolorbox}[promptbox={Prompt 5: Issue Comprehension ExpAgent (Failure)}]
\begin{lstlisting}[basicstyle=\ttfamily\footnotesize, breaklines=true]
You are a bug resolution expert. You will be given a software issue, the corresponding golden patch and a trajectory that represents how an agent attempted to resolve this issue but failed.

## Guidelines
You need to extract some reflections from this failed trajectory according to the golden patch:
1. **reflections** - three reflections on why this trajectory failed to resolve this issue, you need to consider the following aspects:
    - `Perspective`: Explain how should you correctly understand the issue according to the golden patch.
    - `Modification`: If the trajectory correctly identified the modification location, what mistakes were made in actual code modification?

## Important Notes:
    - Reflections should be at the level of thinking, not specific implementation details.
    - Reflections should be expressed in as generalized and abstract terms as possible.
    - Be comprehensive and detailed as possible.
    - Do not include specific object names in the output.

Your output must strictly follow the JSON format shown below:
{
    "perspective": [
        "<one key reflection>",
        ...
        ],
    "modification": [
        "<one key reflection>",
        ...
        ]
}
\end{lstlisting}
\end{tcolorbox}

\subsection{Modification ExpAgent}

\begin{tcolorbox}[promptbox={Prompt 6: Modification ExpAgent Prompt}]
\begin{lstlisting}[basicstyle=\ttfamily\footnotesize, breaklines=true]
You are a software patch refinement expert. You will be given a software issue, a successful trajectory that shows how the agent modified the code to fix the bug, and the agent-generated patch which successfully resolved this issue.

Your job is to: 
1. Compare the generated patch with the issue, determine why this patch could resolve this issue and how to resolve this kind of issue.
2. Analyze the successful trajectory and decide which code modification is vital to resolve this issue.

## Guidelines
Your need to extract and summarize one key insight based on the agent's successful patch:
1. **experience** – abstract the reasoning behind this code change. What principle, pattern, or insight can be generalized from this fix and applied to future debugging cases?

## Important Notes:
    - experience explains *why* the fix worked, in abstract and transferable terms.
    - You could extract *at most three* experiences.
    - Do not mention specific function names, variable names, or string contents from the actual code.

## Output Format
Your output must strictly follow the JSON format shown below:
{      
    "modification": {
        "experience": [
            "<1–2 sentences summarizing the abstract insights learned from making this fix.>",
            ...
        ]
    }
}
\end{lstlisting}
\end{tcolorbox}

\subsection{RerankAgent}

\begin{tcolorbox}[promptbox={Prompt 7: RerankAgent Prompt}]
\begin{lstlisting}[basicstyle=\ttfamily\footnotesize, breaklines=true]
You are a knowledgeable issue resolution assistant. Your task is to analyze a current issue and identify the most relevant past experience that can help resolve it.

You will be given:
- A `problem_statement` describing the current issue
- A set of past trajectories, each with:
  - `issue_id`: A unique identifier
  - `issue_description`: The description of the past issue
  - `experience`: Either a `perspective` (how this successful trajectory understood this issue) or `reflections` (insights gained from an unsuccessful trajectory)

Your job is to:
1. Compare the current `problem_statement` with each past trajectory's `issue_description` and `experience`.
2. Select up to **{k}** past experiences — choose only those that are clearly relevant and potentially helpful for resolving the current issue.
3. You must select **at least one** experience, even if fewer than {k} are strongly relevant.

You should **prioritize trajectories whose problem-solving approach (as described in the perspective) aligns closely with the current issue**.

You must output a JSON object with exactly two fields for each selection:
- `issue_id`: ID of the past issue
- `reason`: A short explanation of why this issue and experience was selected

Your output must strictly follow the JSON format below:
{{
    "issue_id": {{
        "reason": "<why you select this issue and corresponding experience>"
    }},
    ...
}}
\end{lstlisting}
\end{tcolorbox}

\subsection{Reuser}

\subsubsection{Reuse Comprehension Experience Prompt}
\begin{tcolorbox}[promptbox={Prompt 8: Reuser – Reuse Comprehension Experience Prompt}]
\begin{lstlisting}[basicstyle=\ttfamily\footnotesize, breaklines=true]
You are a knowledgeable issue resolution assistant. Your task is to analyze a current issue and generalize the received experiences into a new insight that is applicable to this issue.

You will be given:
- A `problem_statement` describing the current issue
- A past trajectory with:
  - `issue_description`: The description of the past issue
  - `experience`: Either a `perspective` (how this successful trajectory understood this issue) or `reflections` (insights gained from an unsuccessful trajectory)

Your job is to:
1. Compare the current `problem_statement` with each past trajectory's `issue_description` and `experience`.
2. Adapt the old experience to the current issue and produce a new applicable experience.
3. Identify the most likely entry point in the codebase - based on the problem statement - that is critical to resolving the current issue.

You must output a JSON object with exactly one field:
- `new_experience`: A new experience statement tailored to the current issue, based on the old experience. **The more detailed the better**

Your output must strictly follow the JSON format below:
{
    "new_experience": "<the new experience>"
}
\end{lstlisting}
\end{tcolorbox}

\subsubsection{Reuse Modification Experience Prompt}
\begin{tcolorbox}[promptbox={Prompt 9: Reuser – Reuse Modification Experience Prompt}]
\begin{lstlisting}[basicstyle=\ttfamily\footnotesize, breaklines=true]
You are a strategic assistant helping an agent improve its next-step instruction in a debugging task.

You are given:
- A `problem_statement`: a natural language description of the current software problem
- A `current_code_exploration_history`: The recent exploration steps taken to understand or debug the current codebase. This may include what has been examined, eliminated, or hypothesized so far.
- An `instruction`: the next step the agent is expected to take
- A list of `experiences`: each offering past insights about how to better approach the corresponding issue.

Your task is to:
1. Analyze how the current `instruction` relates to the given `issue` and `current_code_exploration_history`
2. Identify useful, transferable, generalized insights from the past experiences of **modification** type
3. Based on those insights, rewrite the instruction to make it more robust, strategically informed, and better suited to succeed in this situation

### Important Notes
    - Focus only on experience of **modification**, and ensure the improved instruction aligns with the original goal but incorporates better reasoning or coverage
    - NEVER add the content that are not related to solving the current problem
    
Output only the following JSON structure:
{
  "enhanced_instruction": "<A single improved and robust instruction, rewritten based on relevant experience of modification type>"
}
\end{lstlisting}
\end{tcolorbox}

\end{document}